\documentclass[11pt]{amsart}
\usepackage{amsmath,amscd,amssymb,amsfonts,amsthm,stmaryrd,color,mathtools}
\usepackage{extarrows}

\usepackage{pdfsync}
\usepackage{hyperref}

\usepackage[cmtip,matrix,arrow,color,poly]{xy}

\newcommand{\cbrack}[2]{\llbracket #1, #2 \rrbracket}

\newtheorem{theorem}{Theorem}[section]

\newtheorem{proposition}[theorem]{Proposition}

\newtheorem{claim}[theorem]{Claim}

\theoremstyle{definition}
\newtheorem{definition}[theorem]{Definition}
\theoremstyle{remark}

\newtheorem{remark}[theorem]{Remark}

\newtheorem{example}[theorem]{Example}

\newcommand{\cC}{\mathcal{C}}
\newcommand{\cD}{\mathcal{D}}

\newcommand{\cK}{\mathcal{K}}
\newcommand{\cL}{\mathcal{L}}
\newcommand{\cM}{\mathcal{M}}
\newcommand{\cN}{\mathcal{N}}
\newcommand{\cO}{\mathcal{O}}

\newcommand{\cY}{\mathcal{Y}}
\newcommand{\cZ}{\mathcal{Z}}

\newcommand{\g}{\mathfrak{g}}

\newcommand{\fX}{\mathfrak{X}}

\newcommand{\bA}{\mathbb{A}}
\newcommand{\bB}{\mathbb{B}}
\newcommand{\bN}{\mathbb{N}}
\newcommand{\bP}{\mathbb{P}}
\newcommand{\bR}{\mathbb{R}}
\newcommand{\bT}{\mathbb{T}}
\newcommand{\bV}{\mathbb{V}}
\newcommand{\bY}{\mathbb{Y}}
\newcommand{\bX}{\mathbb{X}}
\newcommand{\bZ}{\mathbb{Z}}

\newcommand{\nX}{\mathbf{X}}
\newcommand{\nA}{\mathbf{A}}
\newcommand{\nB}{\mathbf{B}}
\newcommand{\nP}{\mathbf{P}}

\newcommand{\Alie}{(A\to M,[\cdot,\cdot],\rho)}

\newcommand{\Cour}[1]      {[\![#1]\!]}

\DeclareMathOperator{\id}{Id}

\DeclareMathOperator{\Lie}{L} 
 
\DeclareMathOperator{\sym}{S^\bullet}

\def\F{\mathfrak{F}}
\def\L{\mathfrak{L}}
\def\Z{\mathbb{Z}}
\def\M{\mathcal{M}}
\def\N{\mathcal{N}}
\def\fun{C}
\def\fC{\mathfrak{C}}

\let\oldtocsection=\tocsection

\let\oldtocsubsection=\tocsubsection

\let\oldtocsubsubsection=\tocsubsubsection

\renewcommand{\tocsection}[2]{\bf\hspace{0em}\oldtocsection{#1}{#2}}
\renewcommand{\tocsubsection}[2]{\hspace{1em}\oldtocsubsection{#1}{#2}}
\renewcommand{\tocsubsubsection}[2]{\hspace{2em}\oldtocsubsubsection{#1}{#2}}

\makeatletter
\@namedef{subjclassname@1991}{2020 Mathematics Subject Classification}
\makeatother


\begin{document}
\sloppy
\title[Dimensional reduction of CSM and Lie theory of Poisson groupoids]{Dimensional reduction of Courant Sigma models and Lie theory of Poisson groupoids}

\author{ Alejandro Cabrera}
\address{Departamento de Matem\'{a}tica Aplicada - IM, Universidade Federal do Rio de Janeiro,
CEP 21941-909, Rio de Janeiro, Brazil}
\email{alejandro@matematica.ufrj.br}

\author{Miquel Cueca}
\address{Mathematics Institute\\Georg-August-University of G\"ottingen\\Bunsenstra{\ss}e 3-5\\G\"ottingen 37073\\Germany}

\email{miquel.cuecaten@mathematik.uni-goettingen.de. \thanks{Corresponding author.}}

\date{\today}

\begin{abstract}
	We show that the 2d Poisson Sigma Model on a Poisson groupoid arises as an effective theory of the 3d Courant Sigma Model associated to the double of the underlying Lie bialgebroid. This field-theoretic result follows from a Lie-theoretic one involving a coisotropic reduction of the odd cotangent bundle by a generalized  space of algebroid paths. We also provide several examples, including the case of symplectic groupoids in which we relate the symplectic realization construction of Crainic-Marcut to a particular gauge fixing of the 3d theory.
\end{abstract}

\maketitle

\tableofcontents

\section{Introduction}

Topological sigma models can incorporate geometric structures, both from the source and target manifolds, and their quantization has shown to be able to provide very interesting results when read in terms of the geometric inputs. Paradigmatic examples include the 2d Poisson Sigma Model and Kontsevich's quantization of Poisson manifolds \cite{cat:kon} and 3d Chern-Simons theory and knot invariants. In this context, the Courant Sigma Model \cite{roy:aksz} appears as a natural generalization of 3d Chern-Simons.

In this paper, we focus on the target space geometry and find a non-trivial relation between instances of the Courant sigma model (CSM) and the Poisson sigma model (PSM). More precisely, we study 3-dimensional Courant sigma models associated to a particular class of Courant algebroids $E=A\oplus A^*$ arising as doubles of Lie bialgebroids $(A,A^*)$. We show that, when the source is a product $\Sigma^{(2d)} \times [0,1]$, this CSM induces an effective theory that can be identified with the 2-dimensional Poisson sigma model on $\Sigma$ and with target being the Poisson groupoid $(G_A,\pi_G)$ integrating $(A,A^*)$ in the Lie-theoretic sense \cite{mac:int}. We also prove an intermediate Lie-theoretic argument needed to connect the infinitesimal information encoded in $(A,A^*)$ with the groupoid one $(G_A,\pi_G)$ in a way that it is compatible with the sigma models formulation.

We begin this introduction by describing the main ingredients involved in these constructions. We then proceed to describe the main Field-theoretic claim and to outline the underlying arguments.

\medskip

\noindent{\bf Main ingredients.} 
A Lie bialgebroid $(A,A^*)$ is given by two Lie algebroid structures, one on the vector bundle $A\to M$ and another on the dual $A^*\to M$, subject to a compatibility condition \cite{mac:bia}. These generalize the familiar Lie bialgebras $(\g,\g^*)$ coming from the theory of quantum groups and Poisson-Lie groups. In the general case, the \emph{double} $E=A\oplus A^*$ inherits the structure of a Courant algebroid in which $A$ and $A^*$ sit as transverse Dirac structures, see \cite{lwx:cou}.

The Lie algebroid structure on $A^*$ induces a linear Poisson structure $\pi_A$ on the total space of $A$ which is \emph{infinitesimally multiplicative} (see \cite{buca:im}). When the Lie algebroid $A$ is integrable by a Lie groupoid, it was shown in \cite{mac:int} that there exists a source-simply-connected integration $G_A\rightrightarrows M$ (unique up to isomorphism) which becomes a \emph{Poisson groupoid}, $(G_A\rightrightarrows M, \pi_{G})$, where $\pi_{G}$ is a \emph{multiplicative} Poisson structure on $G_A$ integrating the linear one $\pi_A$. 

%

\smallskip

\noindent{\bf Main Field-theoretic claim:}
 Let $(A,A^*)$ be a Lie bialgebroid with the double Courant algebroid $E=A\oplus A^*$. Denote by  $\Sigma$ a closed oriented surface and $I=[0,1]$. Assume that $A$ is an integrable Lie algebroid and denote $(G_A\rightrightarrows M, \pi_G)$ the corresponding Poisson groupoid.

\begin{claim}[Main field-theoretic result]\label{main} The Courant sigma model (CSM) on the cylinder $\Sigma\times I$ with target  $E=A\oplus A^*$ and boundary conditions determined by  $A^*\subset E$  has the Poisson sigma model (PSM) on $\Sigma$ with target $(G_A,\pi_{G})$ as an effective theory:
for an appropriate subset of observables $\cO$ and gauge fixings of the CSM,
\begin{equation*}
\langle \cO\rangle_{CSM(\Sigma\times I, E)}=\langle \cO_{red}\rangle_{PSM(\Sigma, G_A)},
\end{equation*}
where $\cO_{red}$ denotes an induced observable in the PSM.
\end{claim}

\noindent The subsets of observables and gauge fixings mentioned in the claim are described in Section \ref{subsub:fixandobs}, as well as the description of the induced observables $\cO_{red}$. We also remark that the claim can be extended to the case when $\partial \Sigma \neq \emptyset$, as explained in Remarks \ref{rmk:corners1}, \ref{rmk:corners2} and \ref{rmk:corners3}.


\smallskip

\noindent{\bf Summary of arguments and outline:} Let us now explain the arguments behind the main claim above and indicate where to find them in the paper.

We begin with preeliminaries. The relation between (target space) supergeometry and underlying ordinary geometric structures is recalled in Section \ref{sec:prelG}, where we also recall the Lie theory for Poisson groupoids and Lie bialgebroids. The overall BV-formalism used for the topological sigma models of this paper is recalled in Section \ref{sec:prelFT}. In particular, the Courant sigma model of our claim is built using the target space $\cM=T^*[2]A[1]$ with structure corresponding to the double of a bialgebroid $(A,A^*)$. 

Next, the first step towards Claim \ref{main} is to observe that there is an exponential map identification
$$ \F\equiv\F_{\Sigma\times I}(\cM):=Map(T[1]\Sigma \times T[1]I, \cM)\overset{(1)}{\simeq} Map(T[1]\Sigma,Map(T[1]I,\cM))=Map(T[1]\Sigma,\cZ)$$
where $\cZ:=Map(T[1]I,\cM)=\F_I(\cM)$. Details are provided in Section \ref{S3}. The boundary conditions are defined by the Dirac structure $A^*$ as 
$$\{\phi\in Map(T[1]I, T^*[2]A[1])\ | \ \phi|_{t=0,1}\in A^*[1]\}.$$

We now consider the relevant field theoretic computations which are of the form
\begin{equation}
\begin{split}
\langle \cO\rangle_{CSM(\Sigma\times I, A\oplus A^*)}:=&\int_{\mathfrak{L}\subset \F_{\Sigma\times I}(\cM) }\sqrt{\mu}\ \cO\ e^{\frac{i}{\hslash}S_{\F_{\Sigma\times I}(\cM)}} \\
\overset{(1)}{=}&\int_{\mathfrak{L}'\subset \F_\Sigma(\F_I(\cM)) }\sqrt{\mu'}\ \cO' \ e^{\frac{i}{\hslash}S_{\F_\Sigma(\F_I(\cM))}}=\int_{\mathfrak{L'}\subset \F_\Sigma(\cZ) }\sqrt{\mu'}\ \cO' \ e^{\frac{i}{\hslash}S_{\F_\Sigma(\cZ)}}\\
\overset{(2)}{=}&\int_{\mathfrak{L}_{red}\subset \F_\Sigma(\cZ_{red})}\sqrt{\mu_{red}}\ \cO_{red}\  e^{\frac{i}{\hslash}S_{\F_\Sigma(\cZ_{red})}}\overset{(3)}{=}\langle \cO_{red}\rangle_{PSM(\Sigma, G_A)}
\end{split}\label{eq:maincompu}
\end{equation}


Step $(1)$ is the exponential identification already mentioned. In Step $(2)$, one recognizes a special structure on $\cZ$. It consists of the existence of an additional (ghost-antighost) grading ``$ga(\cdot)$" and an underlying coisotropic submanifold $\cC$ inside of its degree zero part $\cZ_{ga=0}$.
This situation was studied in \cite{ale:red} where conditions on $\mu'$, $\cL'$ and $\cO'$ where identified for the computation to (formally) descend to the symplectic reduction $\cZ_{red} = \cZ_{ga=0}//\cC$. This is recalled in Section \ref{sec:redaksz}.

Step $(3)$ involves a Lie-theoretic result. We find that the manifold $\cZ_{ga=0}$ yields $T^*[1]Map(I,A)$, the odd cotangent bundle to the space of paths on $A$, while $\cC\subset \cZ_{ga=0}$ is induced by the subspace of \emph{algebroid paths}. Extending the integration picture in which $G_A$ is identified with algebroid paths modulo algebroid homotopies (see \cite{cat:poi},\cite{crfer}), we show that, due to our choice of boundary conditions, the corresponding symplectic reduction yields 
$$\cZ_{red}=T^*[1]G_A, \quad S_{\F_\Sigma(\cZ_{red})} = S_{\F_\Sigma(T^*[1]G_A)}. $$
These results are explained in Section \ref{sec-red}, with \ref{subsec:rlie} devoted to the Lie-theoretic constructions only involving paths and \ref{subsec:rft} detailing how to apply it to obtain Claim \ref{main}.

Altogether, we obtain that the PSM on $(G_A,\pi_G)$ arises as an effective theory (see Section \ref{subsec:BV}) for the CSM after ``integrating out" some of the involved fields. Finally, some examples are discussed in Section \ref{sec:examples}, including the use of a Poisson spray as a gauge fixing and the relation to the symplectic realization of \cite{cra:on}.

\bigskip

We finish the Introduction commenting on known particular cases, providing an outlook of possible developments and commenting on the style of presentation.

\smallskip

\noindent{\bf Related results in the literature.} In \cite{cat:2d} it was studied a particular case of the above claim, which serves as an inspiration for our work. It consists of the case in which the Lie algebroid structure on $A$ is trivial, $\rho=0,[\cdot,\cdot]_A=0$, thus having only a nontrivial dual Lie algebroid structure on $A^*$. The case considered in \cite{cat:2d} can be taken to be a ``linear version'' of our non-linear results: in that case, $G_A\rightrightarrows M$ is given by $A\to M$ seen as a groupoid with fiberwise addition.

On another direction, similar effective ``dimensional reduction" from $\Sigma \times I$ to $\Sigma$ have been largely studied. In most cases, though, the resulting theory on $\Sigma$ is conformal instead of topological as in our case above (see the cases of Chern-Simons, i.e. when $M$ is a point, and WZW, see e.g. \cite{mnev:hol, wit:qua}). Topological boundary conditions for Abelian Chern-Simons were studied in \cite{kap:top} and a mixture of topological with non-topological boundaries appear recently in \cite{sev:ren, sev:plt}. It would be interesting to relate these cases to our case. We also point that the usual ``dimensional reduction" from $\Sigma\times S^1$ to $\Sigma$ have been studied in the Chern-Simons case see e.g. \cite{bla:der}.

Finally, the Hamilton-Jacobi action for AKSZ theories was studied in the recent work \cite{cat:ham}. On cylinders, this action was shown to give rise to a boundary theory which, in turn, is the leading order of the effective action for a special choice of gauge-fixing in the original theory. Our result suggests that, for the CSM on a cylinder $\Sigma\times I$ and with appropiate boundary conditions, the corresponding Hamilton-Jacobi action might coincide with that of the PSM on $\Sigma$. This will be explored elsewhere.

\medskip

\noindent{\bf Outlook.} The main possible application of the heuristic Field-theoretic manipulations is to obtain nontrivial quantum computations. An instance would be the use of Chern-Simons with source $\Sigma \times I$ and target $\mathfrak{g}\bowtie\mathfrak{g}^*$ to make interesting computations on the PSM with source $\Sigma$ and target the Poisson-Lie group $(G,\pi_G)$, or vice-versa. This will be studied elsewhere. See also \cite{cat:ham, cat:cyl} for recent developments.

\medskip

\noindent{\bf About the presentation of the paper.} In this paper, we will focus on providing global geometric constructions for all the ingredients needed. Once these global definitions are established, we can then proceed to verify some of their properties through simple local coordinate computations. This also helps to make the connection with familiar expressions for the field theories present in the literature.

\subsection*{Acknowledgments} This project benefit from multiple conversations with Francesco Bonechi and Alberto Cattaneo. We also thank the anonymous referee for suggestions that improved the paper. M.C. thanks the hospitality of INFN Sezione di Firenze and Universit\"{a}t Z\"{u}rich during two short visits. 
A.C. was supported by the grants FAPERJ E26/203.262/2017, CNPq 305850/2018-0, CNPq 429879/2018-0 and CNPq 309847/2021-4.

\subsection*{Data availability statement}
Data sharing not applicable to this article as no datasets were generated or analysed during the current study.

\section{Preliminaries I: relevant geometric structures and supergeometry}\label{sec:prelG}

In this section, we begin recalling Lie bialgebroids, their doubles given by Courant algebroids and their Lie-theoretic correspondence with Poisson groupoids. In the last subsection, we also recall how these structures are encoded as various types of tensors on supermanifolds. This last presentation is the one which makes the connection with the topological field theories we study in the rest of the paper.

\subsection{Lie bialgebroids and Courant algebroids}\label{S.bia}

A common generalization of both Lie algebras and integrable distributions is given by the concept of a {\bf Lie algebroid} $\Alie$, which also plays a key role in Poisson geometry. It consists of a vector bundle $A\to M$ endowed with a vector bundle morphism $\rho:A\to TM$, called the anchor, and a Lie bracket on the space of sections $[\cdot,\cdot]:\Gamma A\times \Gamma A\to \Gamma A$ satisfying 
$$ [a,fb]= f[a,b]+ (L_{\rho(a)}f) \ b, \ a,b \in \Gamma A, f \in C^\infty(M).$$
Given a Lie algebroid structure $(A\to M, [\cdot,\cdot],\rho)$, the total space $A^*$ of its dual vector bundle $A^* \to M$ inherits a (fiberwise) linear Poisson structure denoted $\pi_{A^*}$ (see \cite[Section 10.3]{mac:book}). Also, the graded algebra of forms $\Gamma \Lambda^\bullet A^*$ inherits a Chevalley-Eilenberg differential $d\equiv d_A$ from the algebroid structure on $A$ and there is an induced Lie derivative operation $L_a=i_a d + d i_a$ for each section $a\in \Gamma A$ (here $i_a$ denotes contraction).

\begin{example}\label{ex-Lie-alg}
The basic examples of Lie algebroids that we will use include:
\begin{enumerate}
	\item Lie algebras $(\g, [\cdot,\cdot])$. In this case, $M=pt$, therefore the anchor is zero. The linear Poisson structure $\pi_{\g^*}\equiv c^{ij}_k x^k \partial_{x^i}\wedge \partial_{x^j}$ on $\g^*$ is the standard one induced by the Lie algebra with $[e^i,e^j]=c^{ij}_k e^k$ for a basis $(e^i)$ of $\g$.
	\item Tangent bundles $(A=TM\to M ,[\cdot,\cdot], \rho=\id)$. The bracket is the usual bracket of vector fields and the anchor is just the identity. The linear Poisson structure $\pi_{T^*M} = \omega_c^{-1}$ is just the standard symplectic one on $A^*=T^*M$.
	\item Cotangent of Poisson manifolds $T_\pi^*M=(T^*M\to M, [\cdot,\cdot]_\pi, \rho=\pi^\sharp)$, with $\pi\in \fX^2(M)$ a Poisson structure on $M$. For $\alpha,\beta\in\Omega^1(M),$ the bracket is given by the formula $[\alpha,\beta]_\pi=L_{\pi^\sharp(\alpha)}\beta-L_{\pi^\sharp(\beta)}\alpha-d\pi(\alpha,\beta)$ and the anchor is $\rho(\alpha)=\pi^\sharp(\alpha)$. In this case, the linear Poisson structure $\pi_{TM}$ is the tangent lift of $\pi$ to $TM$ (see \cite[Section 10.1]{mac:book}).
\end{enumerate}
\end{example}

\noindent A {\bf Lie bialgebroid}, denoted $(A,A^*)$ consists of Lie algebroid structures on both $A\to M$ and its dual $A^*\to M$ satisfying a suitable compatibility condition. One way of stating this condition is that the linear Poisson structure $\pi_A$ (on the total space of $A$) coming from the Lie algebroid on $A^*$ is \emph{infinitesimally multiplicative} with respect to $(A,[\cdot,\cdot],\rho)$, see \cite{buca:im}. We will see below (Section \ref{subsec:super}) a more explicit formulation of the compatibility in terms of supergeometry which is better adapted to the purposes of this paper. For the moment, we observe that the notion is self-dual: $(A,A^*)$ defines a Lie bialgebroid if and only if $(A^*,A)$ does.

Before moving forward, we mention two important particular examples of Lie bialgebroids. In the case where $M=pt$ is just a point, $(A=\g,A^*=\g^*)$ reduces to an ordinary Lie bialgebra. When $(M,\pi)$ is an ordinary Poisson manifold, the induced cotangent Lie algebroid $A=T^*_\pi M$ is endowed with a natural Lie bialgebroid structure for which $A^*=TM$ reduces to the standard tangent algebroid \cite{mac:bia}. In this last case, as in the above examples, the canonical symplectic $\pi_A=\omega_c^{-1}$ is the one that becomes infinitesimally multiplicative for $A=T^*_\pi M$.

We now proceed to describe Courant algebroids which appear as doubles for such $(A,A^*)$. As a motivation, we recall that, in the case of Lie bialgebras $(\g,\g^*)$, the Drinfeld double structure on $\g\oplus\g^*$ (which is again a Lie bialgebra) plays an important role in the theory. For general Lie bialgebroids, the corresponding doubles were studied in \cite{lwx:cou} where it is shown that it leads to the new notion of a {\bf Courant algebroid}. A Courant algebroid $(E\to M, \langle\cdot,\cdot\rangle, \Cour{\cdot,\cdot}, a)$ is a vector bundle $E\to M$ endowed with a non-degerate symmetric pairing $\langle\cdot,\cdot\rangle$, an anchor $a: E\to TM$ and a bracket $\Cour{\cdot,\cdot}:\Gamma E\times\Gamma E\to \Gamma E$ satisfying:
\begin{itemize}
\item $\cbrack{e_1}{fe_2} = a(e_1)(f) e_2 + f\cbrack{e_1}{e_2}$,
\item $a(e_1)(\langle e_2,e_3\rangle) = \langle\cbrack{e_1}{e_2},e_3\rangle + \langle e_2,\cbrack{e_1}{e_3}\rangle$,
\item $\cbrack{\cbrack{e_1}{e_2}}{e_3} = \cbrack{e_1}{\cbrack{e_2}{e_3}} - \cbrack{e_2}{\cbrack{e_1}{e_3}}$,
\item $\cbrack{e_1}{e_2} + \cbrack{e_2}{e_1} = \cD \langle e_1,e_2\rangle$,
\end{itemize}
 where $\cD : C^\infty(M) \to \Gamma(E)$ is defined by $\langle\cD f, e\rangle = a(e)(f).$
 
 In the case of a Lie bialgebroid $(A,A^*)$ the double is given by $E=A\oplus A^*\to M$ with the following Courant algebroid structure
 \begin{equation}\label{double}
 \begin{split}
 &E=A\oplus A^*\to M, \quad \langle e_1, e_2\rangle= \langle b_1, a_2\rangle+\langle b_2, a_1\rangle,\quad a(e_1)=\rho(b_1)+\widetilde{\rho}(a_1),\\
& \Cour{e_1, e_2}=[b_1, b_2]+\Lie^*_{a_1}b_2-i_{a_2}d^*b_1+[a_1, a_2]_*+\Lie_{b_1}a_2-i_{b_2}da_1,
 \end{split}
 \end{equation}
where $e_i=b_i+a_i$ with $b_i\in\Gamma A, \ a_i\in\Gamma A^*$ for $i=1,2$, $\rho$ (resp. $\tilde \rho$) is the anchor of $A$ (resp. $A^*$) and $\Lie, d$ (resp. $\Lie^*, d^*)$ comes from the Cartan calculus of $A$ (resp. $A^*$) recalled above.

\subsection{Lie theory for bialgebroids and path integration}\label{subsec:lie}
In this subsection, we recall the following ingredients of Lie theory: differentiation of Lie groups into Lie algebras can be generalized into
\begin{eqnarray*}
\text{Lie groupoids} &\overset{\text{Lie functor}}{\to}& \text{Lie algebroids}\\
(G\rightrightarrows M) &\mapsto& (A_G\to M),
\end{eqnarray*} 
and that it can be refined to yield a correspondence
$$ \text{Poisson groupoids, $(G\rightrightarrows M, \pi_G)$} \to \text{Lie bialgebroids, $(A,A^*)$}.$$
In this way, Lie bialgebroids are seen as infinitesimal counterparts of so-called Poisson groupoids and, at the end, we describe some ingredients entering the converse ("integration") constructions involving algebroid paths.

We begin observing that, just as Lie algebras are the infinitesimal versions of Lie groups, Lie algebroids can be seen as infinitesimal versions of Lie groupoids. Recall that a groupoid $G\rightrightarrows M$ can be described as a category in which every morphism is invertible, $M$ denoting the set of objects and $G$ the set of morphisms or arrows. When $M$ and $G$ are manifolds\footnote{It is customary to allow $G$ to be non-Hausdorff, though the $s$-fibers must be.} and the source and target maps $s,t:G\to M$ are submersions, we say that $G\rightrightarrows M$ defines a {\bf Lie groupoid} (see \cite{mac:book} for an extensive treatment). In particular, for each object $x\in M$, there is one identity $1_x\in G$ and composition of arrows is only partially defined on $G$, namely, $m(g,h)\in G$ for $g,h\in G$ is defined when $s(g)=t(h)$.

Ordinary Lie theory can be extended to a large extent to the context of Lie algebroids and Lie groupoids. The above mentioned Lie functor takes a Lie groupoid $G\rightrightarrows M$ and associates a Lie algebroid $A_G$ given by $A_G=\ker Ts_{|M}\to M,$ anchor $\rho= Tt|:A_G\to TM$ and bracket induced by the Lie bracket of right-invariant vector fields. In this sense, elements in the algebroid $A_G$ can be seen as infinitesimal arrows deforming the identity ones with their source being kept fixed. The standard Lie theorems I and II, which state that the infinitesimal information uniquely characterizes a given groupoid morphism up to (source fibers) coverings still hold true. Nevertheless, unlike Lie III in the usual theory, not every Lie algebroid $\Alie$ is isomorphic to $A_G$ for some Lie groupoid $G\rightrightarrows M$, see \cite{crfer} where the corresponding theory of obstructions is developed. When it is the case, we say that the Lie algebroid is {\bf integrable} and we denote by $G_A\rightrightarrows M$ the integration of $A$ with $1$-connected source fibers (which is unique up to isomorphism), so that $A_{G_A}\simeq A$.

\smallskip

We now proceed to describe the refinement of this Lie theory targeting Lie bialgebroids $(A,A^*)$. As it turns out, when $A$ is integrable, $G_A$ inherits an extra Poisson structure $\pi_G$ coming from the linear one $\pi_A$ on $A$ (which, in turn, is defined by the dual algebroid on $A^*$ as recalled in Section \ref{S.bia}). The resulting structure is axiomatized as that of a {\bf Poisson groupoid}: it consists of a pair $(G\rightrightarrows M, \pi_G)$ where $\pi_G$ defines a Poisson structure on the space of arrows $G$ such that $graph(m)\subseteq G\times G\times \bar{G}$ is a coisotropic submanifold, the bar means that we consider $-\pi$ instead of $\pi$. Structures on $G$ satisfying this condition are called \emph{multiplicative}. Poisson groupoids were introduced in \cite{Wei:coi} in order to unify Poisson Lie groups and symplectic groupoids. 

Given a Poisson groupoid $(G\rightrightarrows M,\pi_G)$, one can verify that the Lie algebroid $A=A_G$ inherits a Lie bialgebroid structure from $\pi_G$, see \cite{mac:bia}. Indeed, $\pi_G$ can be differentiated to a linear Poisson structure $\pi_A$ on $A$, thus defining an algebroid on $A^*$, and which is infinitesimally multiplicative as recalled above (see also \cite{buca:im}). In \cite{mac:int} it was show the converse integration result:
\begin{theorem}[Mackenzie-Xu \cite{mac:int}]
	Let $(A,A^*)$ be a Lie bialgebroid and assume that $A$ is integrable. Then, $G_A$ inherits a unique Poisson groupoid structure $(G_A\rightrightarrows M, \pi_G)$ whose differentiation yields $(A,A^*)$. 
\end{theorem}
In the case of a Lie bialgebra $(\g,\g^*)$, the group $(G_A\equiv G, \pi_G)$ becomes a familiar Poisson-Lie group and the integration result was proved by Drinfeld. The Lie bialgebroid $(T^*_\pi M, TM)$ associated to an ordinary Poisson manifold $(M,\pi)$ yields, in the integrable case, a Poisson groupoid $(G_A,\pi_G)$ in which $\pi_G$ is symplectic, establishing the well-known correspondence between Poisson manifolds and symplectic groupoids (see more details in e.g. \cite{mac:int}).

\medskip

\noindent{\bf Integration through paths.} We finish this subsection by recalling the following concrete construction of $G_A$ following \cite{cat:poi, crfer}. Given a Lie algebroid $\Alie$, one considers the space of {\bf algebroid paths} $PA$ consisting of vector bundle morphisms $TI\to A, \ (t,\partial_t)\mapsto a(t)$ ($I=[0,1]$) satisfying the condition 
\begin{equation}\label{path-eq}
\frac{d}{dt}x(t) = \rho(a(t)), \text{ where $x(t)\in M$ is the projection of $a(t)$ along $A\to M$.}
\end{equation}
There is a corresponding notion of algebroid homotopy (with fixed end-points) defined by algebroid morphisms $T(I\times I) \to A$ subject to suitable boundary conditions and inducing an equivalence relation $\sim$ on $PA$. With these ingredients, one obtains
\begin{equation}
G_A = PA/\sim
\end{equation}
which can be endowed with a groupoid structure over $M$ coming from concatenation of paths. In \cite[Section 2.1]{crfer} it is shown that, when $A$ is integrable, $PA/\sim$ inherits a smooth structure turning it into the integration $G_A\rightrightarrows M$ of $A$ introduced above. When $A=T^*_\pi M$ comes from a Poisson manifold $(M,\pi)$, the quotient $PA/\sim$ can be interpreted as a symplectic reduction from the space of all paths $Map(I,T^*M)$ and also as the physical phase space of the PSM with source $\Sigma=I\times I$, see \cite{cat:poi}.

\begin{example} (Integrating a cotangent lift \cite{cat:on})\label{ex:cotangpaths}
	Let $A\to M$ be a vector bundle. Then, there is a natural vector bundle structure $T^*A\to A^*$ and an identification $T^*A\simeq T^*A^*$ which brings the latter into the ordinary cotangent structure. When $A$ is a Lie algebroid, $T^*A\to A^*$ inherits a natural Lie algebroid structure over $A^*$ called the {\bf cotangent lift} of $A$. This can be identified as a particular case of Example \ref{ex-Lie-alg}(c) coming from the Poisson structure $\pi_{A^*}$ on $A^*$ (see \cite[Section 9.4]{mac:book}).  When $G_A\rightrightarrows M$ integrates $A$, then $T^*G_A\rightrightarrows A^*$ inherits a Lie groupoid structure\footnote{In fact, is a VB-groupoid over $G_A\rightrightarrows M$, see \cite[Section 11.3]{mac:book} for details.} which integrates $T^*A$, \cite{cat:on, mac:book}.
	 Let us discuss algebroid paths into $(T^*A\to A^*)$ for latter reference. In this case, an algebroid path $TI\to T^*A$ projects, under the natural map $T^*A \to A$, to an algebroid path in $A$. Hence we say that such a $T^*A$-path is a \emph{cotangent lift} of the base $A$-path. For later comparison, let us spell out the $T^*A$-path equations on a trivializing chart $(x^i, a^\alpha, p_i, p_{a^\alpha}=b_\alpha)$ induced by a trivializing chart $(x^i, a^\alpha)$ for $A$. As equation \eqref{path-eq} show, we just need to know the anchor of $T^*A\to A^*$.  Example \ref{ex-Lie-alg}(c) combined with the isomorphism $T^*A\cong T^*A^*$  shows that the anchor is given by $\pi_{A^*}^\sharp$ yielding that, on top of eq. \eqref{path-eq} for $(x^i,a^\alpha)$, a $T^*A$-path must satisfy 
\begin{equation}\label{eq:cotagpath}
	 \frac{d}{dt}\Big( x^i, b_\alpha\Big)(t)=\Big(\rho^i_\alpha a^\alpha,  -\rho^i_\alpha p_i-c_{\alpha\beta}^\gamma b_\gamma a^\beta\Big)(t).
\end{equation}  
\end{example}

For general Lie bialgebroids, the induced $\pi_G$ on $G_A$ can be explicitly described in terms of paths, using the identification $G_A=PA/\sim$ described above, out of the infinitesimal data $\pi_A$ on $A$, see \cite{xu:uni}. (See also \cite{buca:im} for a formulation in terms of algebroid morphisms.) We shall come back to this description in Section \ref{subsec:rlie}.

\subsection{Supergeometric formulation}\label{subsec:super}
In this paper, we make extensive use of the language of supermanifolds and $\bN$-graded supermanifolds. We thus first review some general definitions (see e.g. \cite{fio:book, cat:int, mnev:lec} for a more detailed treatment) and then proceed to recall known descriptions of the Lie bialgebroids and their doubles in supergeometric terms. These descriptions will be used in the sequel.

We begin recalling that a supermanifold (resp. a $\bN$-graded supermanifold) $\cM=(M, C^\infty(\cM))$ is a ringed space where $M$ is a smooth manifold and $C^\infty(\cM)$ is a sheaf of $\bZ_2$-graded (resp. $\bN$-graded) commutative algebras with a local model given by
$$C^\infty(\cM)_{|U}\cong C^\infty(U)\otimes \sym V^*,$$
where $V$ is a purely odd (resp. $\bN_{>0}$-graded) vector space and $\sym$ denotes the graded symmetric algebra.  
On a supermanifold, there is thus an atlas of coordinates $(x^i,\theta^a)$ which split into two subsets, the even (Bosonic) ones $x^ix^j=x^jx^i$ and the odd (Fermionic) ones $\theta^a\theta^b=-\theta^b\theta^a$. The structure of an $\bN$-graded supermarmanifold can be seen as a refinement of the ($\bZ_2$-)supermanifold structure in which there is an atlas by graded coordinates $(x_0^{i_0}, y^{j_n}_{(n)})_{n> 0}, \ |x_{0}^{i_0}|=0, \ |y_{(n)}^{j_n}|=n$, where $|\cdot|$ denotes the degree and the induced $\bZ_2$-parity is given by $n \text{ mod } 2\in \bZ_2$, and changes of coordinates are polynomial for degrees $>0$. It is interesting to recall that this $\bN$-refinement can be globally encoded as a smooth action of the multiplicative monoid $\mathbb{R}$ on the supermanifold $\cM$ (see \cite{gra:graded}). 

\begin{remark}
	We observe that $\bZ$-graded supermanifolds can be defined analogously and that they play an important role in the BV formalism by allowing both positive and negatively graded coordinates. One technical detail to have in mind is that, in general, changes of coordinates can involve (formal) infinite power series in which the total degree of each term is a given fixed $m\in \bZ$. The $\bZ$-graded supermanifolds which play an important role in our description (especially in Section \ref{sec-red}) have a very controlled structure, being defined from underlying vector bundles, and the above possible convergence problems do not appear. 
\end{remark}

\smallskip

Basic elements of differential geometry can be defined on supermanifolds and $\bN$-graded supermanifolds. In particular, (multi-)vector fields $\fX(\cM)$ and differential forms $\Omega(\cM)$ can be defined and super/graded versions of Cartan calculus hold. In the case $\cM$ carries an additional $\bN$-grading, there is a subset of vector fields and forms which are homogeneous of degree $j$ which are denoted by lower indices, i.e. we will have $C_j^\infty(\cM)$ for functions of degree $j$ or $\Omega^k_j(\cM)$ for differential $k$-forms of degree $j$ or $\fX^k_j(\cM)$ for  $k$-multivector fields of degree $j$. We refer to this inherited grading given by $j$ as the ``internal degree" and is denoted by $|\cdot|$.

\begin{example}\label{ex:coordsuper}
	A simple example is given by $\cM= \bR^{n_0} \times \bR^{n_1}[1]$ where the $[1]$ indicates that the latter linear coordinates $a^\alpha$ on $\bR^{n_1}$ are defined to have degree $1$ while the $x^i$ on $\bR^{n_0}$ are standard, degree zero, ones ($M=\bR^{n_0}$ here). In this case, we can provide simple examples of internally-graded tensors on $\cM$ to fix ideas:
	$$ f(x) \in C^\infty(\bR^{n_0}) \subset C_0^\infty(\cM), \ f_{\alpha \beta}(x) a^\alpha a^\beta \in C_2^\infty(\cM),$$
	$$ \omega_{ij}(x) dx^idx^j \in \Omega^2_0(\cM), \ c_{\alpha i \beta}(x) a^\alpha dx^ida^\beta, \omega_{\alpha \beta}(x) da^\alpha da^\beta \in \Omega^2_2(\cM),$$
	$$ \pi^{ij}(x)\partial_{x^i}\partial_{x^j} \in \fX^2_0(\cM), \pi^{\alpha \beta}_\gamma(x) a^\gamma \partial_{a^\alpha}\partial_{a^\beta} \in \fX^2_{-1}(\cM). $$
	(Observe the internal degrees $|a^\alpha|=1=|da^\alpha|$ and, then, $|\partial_{a^\alpha}|=-1$.)
\end{example}

The most important notions on $\bN$-graded supermanifolds $\cM$ that we shall use are the following:
\begin{itemize}
	\item a {\bf $Q$-manifold} $(\cM,Q)$ is given by a vector field $Q \in \fX^1_1(\cM)$ of internal degree $1$ (which thus it is an odd derivation of $C^\infty(\cM)$) which is homological in the sense that $[Q,Q]=2Q^2=0$. The algebra of global sections $C^\infty(\cM)$ becomes a d.g.a. with differential $Q$.
	
	\item a {\bf $QP$-manifold of (internal) degree $n$} $(\cM,\omega,Q,\theta)$ is defined by a symplectic structure $\omega \in \Omega^2_{n}(\cM)$ of internal degree $n$ (which can thus be even or odd depending on the parity of $n$), a vector field $Q \in \fX^1_1(\cM)$ and a function $\theta\in C^\infty_{n+1}(\cM)$ such that $Q$ is the hamiltonian vector field associated to $\theta$,
	$$ i_{Q} \omega = d\theta \in \Omega^1_{n+1}(\cM)$$
	and the \emph{Classical Master Equation} holds,
	$$ \{\theta,\theta\}_\omega = 0,$$
	where $\{,\}_\omega$ denote the graded Poisson brackets induced by $\omega$ (which are of internal degree $-n$).
\end{itemize}
We observe that $Q$ in a $QP$-manifold structure defines a $Q$-manifold one as a consequence of the classical master equation for the hamiltonian $\theta$. Since $Q$ is completely determined by the rest of the data, we often simplify and refer to $(\cM,\omega,\theta)$ as defining the QP-manifold structure. Also, the definitions have a straightforward pure $\bZ_2$-version (just consider $n\in \bZ_2$ above) as well as $\bZ$-graded ones.

\medskip

\noindent{\bf Encoding Lie bialgebroids and their doubles supergeometrically.} Let us go back to the setting of the previous subsections. Given a vector bundle $A\to M$, we can define an $\bN$-graded supermanifold $\cM=A[1]$ by declaring $$C^\infty(\cM)=\Gamma \Lambda^\bullet A^*$$
with its standard grading. This corresponds to a global version of Example \ref{ex:coordsuper} in which sections of $A^*$ are seen as fiberwise linear odd coordinates on $A[1]$ of internal degree 1. The first important result for this article is the following supergeometric  description of Lie algebroids.

\begin{proposition}[Vaintrob \cite{vai:lie}]
	Let $A\to M$ be a vector bundle. The following structures are in $1:1$ correspondence:
	\begin{enumerate}
		\item$\Alie$ Lie algebroid structures on $A\to M$.
		\item $(A[1], Q_A)$ Q-manifold structures on $A[1]$, i.e. $Q_A\in\fX^1_1(A[1])$ with $[Q_A,Q_A]=0$.
		\item $(A^*[1], \pi_{A^*})$ degree $-1$ Poisson structures on $A^*[1]$,  $\pi_{A^*}\in\fX^2_{-1}(A[1])$ with $[\pi_{A^*},\pi_{A^*}]=0$.
	\end{enumerate}
\end{proposition}
Let us review this correspondence using local coordinates. If $\{x^i\}$ are coordinates in $M$, $\{b_\alpha\}$ a basis of sections for $A$, inducing fiber coordinates on $A^*[1]$, and $\{a^\alpha\}$ the dual basis, inducing fiber coordinates on $A[1]$; then a local coordinate description of the preceding structures is given by 
\begin{eqnarray}
\quad&\rho(b_\alpha)(x^i)=\rho^\alpha_i\quad\text{and}\quad [b_\alpha, b_\beta]= c_{\alpha\beta}^\gamma b_\gamma \quad \text{with}\quad \rho^\alpha_i,  c_{\alpha\beta}^\gamma\in C^\infty(M);\\
\quad&\fun^\infty(A[1])\supset \{ x^i, a^\alpha\} \ \text{ with }\ |x^i|=0, |a^\alpha|=1\quad \text{and}\quad Q_A= \rho^i_\alpha a^\alpha \frac{\partial}{\partial x^i}+\frac{1}{2} c^\gamma_{\alpha\beta} a^\alpha a^\beta \frac{\partial}{\partial a^\gamma};\\
\quad&\fun^\infty(A^*[1])\supset \{x^i, b_\alpha\} \ \text{ with }\ |x^i|=0, |b_\alpha|=1\quad \text{and}\quad \pi_{A^*}= \rho^i_\alpha \frac{\partial}{\partial b_\alpha} \frac{\partial}{\partial x^i}+\frac{1}{2} c^\alpha_{\beta\gamma} b_\alpha  \frac{\partial}{\partial b_\beta}\frac{\partial}{\partial b_\gamma}.\label{Poi-coord}
\end{eqnarray}
Notice that, since $\pi_{A^*}$ is linear with respect to the bundle structure, it has the same expression on both the ordinary $A^*$ (as recalled before) and on the shifted $A^*[1]$.

Now, consider a Lie bialgebroid $(A,A^*)$. The algebroid structure on $A$ corresponds to the vector field $Q_A$ on $A[1]$ while the algebroid structure on $A^*$ corresponds to the Poisson structure $\pi_A$ on $A[1]$. The needed properties stating that $(A[1], Q_A, \pi_A)$ defines a Lie bialgebroid can be summarized as follows: $Q_A\in\fX^1_{1}(A[1])$ and $\pi_A\in\fX^2_{-1}(A[1])$ satisfy
\begin{equation}\label{Q-pi}
\Lie_{Q_A} Q_A=[Q_A,Q_A]=0,\quad [\pi_A, \pi_A]=0\quad\text{ and }\quad  \Lie_{Q_A}\pi_A=[Q_A,\pi_A]=0.
\end{equation}
Indeed, using the preceding equivalence, we see that the first two equations say that we have induced Lie algebroids $(A\to M, [\cdot,\cdot],\rho)$ and $(A^*\to M, [\cdot,\cdot]_*,\widetilde{\rho})$, while the last condition is the compatibility between them. It is not hard to see that if $(A[1], Q_A, \pi_A)$ is a Lie bialgebroid then $(A^*[1], Q_{A^*}, \pi_{A^*})$ is also a Lie bialgebroid where $Q_{A^*}\leftrightarrow \pi_A$ and $\pi_{A^*}\leftrightarrow Q_A$. For the classical definition of Lie bialgebroids see  e.g. \cite{mac:bia}, the equivalence to the present description is proven in  \cite{roy:on, vor:gra}.

\begin{remark}
	In coordinates as above, a Lie bialgebroid structure $(A[1], Q_A, \pi_A)$ is given by
	$$\fun^\infty(A[1])\supset \{x^i, a^\alpha\} \quad \text{with}\quad |x^i|=0, \ |a^\alpha|=1,$$
	$$Q_A=\rho^i_\alpha a^\alpha \frac{\partial}{\partial x^i}+\frac{1}{2} c^\gamma_{\alpha\beta} a^\alpha a^\beta \frac{\partial}{\partial a^\gamma} \quad\text{and}\quad \pi_A=\widetilde{\rho}^{\alpha i} \frac{\partial}{\partial a^\alpha} \frac{\partial}{\partial x^i}+\frac{1}{2} \widetilde{c}_\alpha^{\beta\gamma} a^\alpha  \frac{\partial}{\partial a^\beta}\frac{\partial}{\partial a^\gamma}.$$
	where $\rho^i_\alpha, c^\gamma_{\alpha\beta}, \widetilde{\rho}^{\alpha i} , \widetilde{c}_\alpha^{\beta\gamma} \in C^\infty(M)$ satisfying relations induced by the equations  \eqref{Q-pi}.
\end{remark}

Similarly, Courant algebroids also have a supergeometric description that help us to understand them.

\begin{theorem}[Roytenberg \cite{roy:on}]\label{sev-roy-corr}
	There is a one to one correspondence between Courant algebroids and degree 2 QP-manifolds.  
\end{theorem}

In the case of the double $E=A\oplus A^*$ of a Lie bialgebroid $(A,A^*)$, as recalled in Section \ref{S.bia}, the associated QP-manifold is given by a graded cotangent lift as follows. Let $(A[1], Q_A, \pi_A)$ be a Lie bialgebroid then $(T^*[2]A[1],\ \omega,\ \theta=Q_A+\pi_A)$ is a degree $2$ QP-manifold where the symplectic structure $\omega$ is the canonical one (as defined on any shifted cotangent bundle) and the functions are given by the identification 
$$C^\infty_k(T^*[2]A[1])=\bigoplus _{2i+j=k}\fX^i_j(A[1]),\quad \text{hence}\quad \theta=Q_A+\pi_A\in C^\infty_3(T^*[2]A[1]).$$ 
Notice that the symplectic Poisson brackets correspond to the Schouten bracket on multivectors $\fX^\bullet( A[1])$ shifted by $2$. The Lie bialgebroid conditions \eqref{Q-pi} for $(A[1], Q_A, \pi_A)$ are equivalent to $\{\theta,\theta\}=0$ and therefore we have a $QP$-manifold structure. The Courant algebroid \eqref{double} can be recovered by the formulas
\begin{equation*}
\Gamma E=C^\infty_1(T^*[2]A[1]),\ \langle e_1, e_2\rangle=\{e_1, e_2\},\ a(e_1)(f)=\{\{e_1, \theta\}, f\},\
\Cour{e_1,e_2}=\{\{e_1, \theta\}, e_2\}
\end{equation*}  
where $e_1,e_2\in\Gamma E$ and  $\{\cdot,\cdot\}$ is the degree $-2$ Poisson bracket defined by $\omega$. For a coordinate description see Section \ref{gen} below.

\medskip

\begin{remark}
	Another structure on supermanifolds that we will be using is that of integration with respect to Berezinian volumes. We shall not detail these further, the reader can consult e.g. \cite{mnev:lec,sch:the} for an account suited to the uses in this paper. For these operations, the relevant structure is the $\bZ_2$-supermanifold one. We also mention that, in the context of field theories and the BV formalism below, $QP$-manifold structures appear (formally) on infinite dimensional spaces of maps between supermanifolds. As customary, we will use the same notations and shall proceed formally as if these were finite dimensional supermanifolds. Finally, we remind the reader that, typically, operations and structures which involve differentiation can be formalized even in infinite dimensions while operations which involve integration require extra care and have to be analyzed in a case-by-case basis.
\end{remark}


\section{Preliminaries II: Field theoretic constructions}\label{sec:prelFT}

\subsection{BV quantization}\label{subsec:BV}

The so-called BV formalism is a device oriented towards quantization of field theories with generalized gauge symmetries through path integrals with additional (super)fields. It includes topological sigma models and it is the general formalism that we will be dealing with in this paper. We thus provide a short summary below, following the geometric description of \cite{sch:the}, and at the same time fix some general notations.

In the BV setting, we have a space of superfields $\F$ (typically a formal infinite dimensional supermanifold) endowed with an odd symplectic structure $\omega_\F$ and a (formal) Berezinian volume $\mu$ satisfying the compatibility condition encoded in the notion of {\bf SP-structure}, see \cite{sch:the}. The key quantum computations are of the form
\[ \langle \cO \rangle = \int_{\L \subset \F} \sqrt{\mu} \ \cO \ e^{\frac{i}{\hbar} S} \]
where $\L \subset (\F,\omega_\F)$ is a lagrangian submanifold implementing a {\bf gauge fixing} condition, $\sqrt{\mu}$ is a naturally induced Berezinian measure on $\L$, and $\cO$ and $S$ are functions on $\F$ representing an observable and the BV-action, respectively. In the presence of internal $\Z$-grading, it is conventionally assumed that $\omega_\F$ is of degree $-1$ so that the BV-action $S$ is of degree zero.

The key fact, which can be rigorously proven when $\F$ is a finite-dimensional SP-manifold (see \cite{sch:the}), is that the integral $\langle \cO \rangle$ is stable under deformations of the gauge fixing $\L\subset \F$ when the following two conditions are met:
\begin{enumerate}
	\item $S$ satisfies the Quantum Master Equation (QME): $\{S,S\}-2i\hbar \Delta S = 0$
	\item $\cO$ is a Quantum observable: $\{S,\cO\}-i\hbar \Delta \cO = 0$.
\end{enumerate}
Here, $\{\cdot,\cdot\}$ denote the (formal) Poisson brackets induced  by $\omega_\F$ and $\Delta$ is the so-called {\bf BV-laplacian} defined by the SP-structure as $\Delta f = \frac{(-1)^{deg(f)}}{2} div_\mu (X_f)$, the $\mu$-divergence of the hamiltonian vector field $X_f$ in $(\F,\omega_\F)$. It is worth noting that the above two conditions combine to give $\Delta(\cO e^{\frac{i}{\hbar}S})=0$ which is the general requirement for the stability of the integral under deformations of $\L$. It is also interesting to have in mind that $\int_{\L\subset \F}\sqrt{\mu} \ \Delta(f) = 0$ for any $f$ when $\L$ is closed\footnote{Schwartz shows in \cite{sch:the} that these facts are equivalent to the classical Stokes theorem under a transformation which maps $\Delta$ to de Rham differential.}.

The leading order in $\hbar$ in the QME yields the Classical Master Equation,
\[ \{S,S\} = 0.\]
Forgetting about the (formal) measure $\mu$, the triple $(\F,\omega_\F, S)$ thus defines an \emph{odd QP-structure} in the sense of Section \ref{subsec:super}. In most relevant cases for us, $\F$ carries an internal $\bZ$-grading for which $\omega_\F$ is of degree $-1$ and $S$ is of degree zero. This is the standard convention for the BV formalism.

\bigskip

Finally, we recall the notion of an \emph{effective theory} within the BV formalism, see \cite{lose:talk} and \cite[Section 4.7]{mnev:lec}. 
The key idea can be seen in the case in which $\F = \F_1 \times \F_2$ as SP-manifolds. In this case, choosing a lagrangian $\L_2\subset \F_2$, we can define an {\bf effective action} $S_{eff} \in C^\infty(\F_1)$ via
\[ e^{\frac{i}{\hbar}S_{eff}} = \int_{\L_2} \ \sqrt{\mu_2} \ e^{\frac{i}{\hbar}S}.\]
It follows directly that $S_{eff}$ satisfies the QME on $\F_1$,
\[ \Delta_1e^{\frac{i}{\hbar}S_{eff}} =  \int_{\L_2}\sqrt{\mu_2} \ \Delta_1 e^{\frac{i}{\hbar}S} = \int_{\L_2}\sqrt{\mu_2}\ (\Delta_1+\Delta_2) e^{\frac{i}{\hbar}S}  = 0. \]
This construction generalizes to the case in which there is a fibration $\F\to \F_1$ adapted to the SP-structures and one replaces $\int_{\L_2}$ by a suitable pushforward, see \cite[Sec. 4.7]{mnev:lec}. In these cases, we refer to $(\F_1,\omega_{\F_1},\mu_1, S_{eff})$ as an {\bf effective theory induced by $(\F,\omega_{\F},\mu, S)$}.

\subsection{AKSZ construction of the space of superfields}\label{gen}

In this section we recall the AKSZ construction \cite{AKSZ} which yields the information of the space of (BV-)superfields $(\F,\omega_\F,S)$, endowed with a relevant QP-structure, out of source and target supermanifolds endowed with appropriate geometric structures. This construction is, by now, standard material, see e.g. \cite{cat:cla, mnev:lec, roy:aksz, sch:the}. 

Let $\cN$ and $\cM$ be supermanifolds (typically finite dimensional and endowed with an additional $\bN$- or $\bZ$-grading). Consider the (formal) supermanifold of maps\footnote{In categorical terms, we consider the inner-hom in the category of supermanifolds.} $Map(\cN,\cM)$ so that $\cN$ is called the source and $\cM$ is the target. Associated to it we have the evaluation map $ev:\cN\times Map(\cN, \cM)\to \cM$ and the induced pullback operation 
$$ ev^*:\Omega^p_q(\cM)\to \Omega^p_q(\cN\times Map(\cN, \cM)),$$
where $\Omega^p_q(\cM)$ denotes $p$-forms which are of internal-degree $q$ (with respect to the extra $\Z$-grading).
When the manifold $\cN$ has a Berezinian volume $\mu$ of degree $n$, an induced \emph{transgression map} is defined by
\begin{equation*}
 \bT_{\cN}:\Omega^p_q(\cM)\to \Omega^p_{q-n}(Map(\cN, \cM)), \quad \bT_{\cN}(\omega)=\int_\cN \mu \ ev^*\omega.
\end{equation*} 
The case which interests us the most is $\cN = T[1]N$, the odd tangent bundle of an ordinary oriented manifold $N$, possibly with boundary $j:\partial N\to N$, which comes endowed with a natural Berezinian volume $\mu$ (defined by the orientation) and with the de Rham vector field $d_N \in \mathfrak{X}(T[1]N)$ (yielding a Q-structure). In this case, we simplify the notation as $\bT_{T[1]N}\equiv \bT_{N}$.

Next, assume the target is endowed with a QP-structure $(\cM,\omega=d\lambda, \theta)$ of internal degree $k$  in which $\omega \in \Omega^2_k(\M)$ is exact with potential $\lambda$. Denote by $Q=\{\theta,\cdot \}$ the corresponding vector field. Hence, on $Map(T[1]N,\cM)$ we have an induced 2-form, a vector field and a function given by 
$$\omega_{\F}=\bT_N(\omega)\in\Omega^2_{k-n}(\F),\quad Q_\F =\widehat{Q}+\widehat{d_N}\in\fX^1_1(\F), \quad S_{\F}=\bT_N(\theta)+i_{\widehat{d_N}}\bT_N(\lambda)\in C^\infty_{k-n}(\F),$$
where $\widehat{Q}$ and $\widehat{d_N}$ denote the natural lifts of $Q$ and $d_N$ to the space of maps (seen as infinitesimal transformations of the target and the source, respectively).

Denote by $j^!:Map(T[1]N,\cM)\to Map(T[1]\partial N,\cM)$ the natural restriction map induced by the inclusion $j:\partial N\to N$. The key properties of this construction are (see \cite{cat:aksz, cat:cla})
 $$\Lie_{Q_\F}\omega_\F=0,\quad [Q_{\F}, Q_{\F}]=0,\quad  i_{Q_\F}\omega_\F=(-1)^n dS_\F +j^{!*}\bT_{\partial N}(\lambda).$$
 We thus see that, when $\partial N=\emptyset$, the space $\F=Map(T[1]N,\M)$ inherits a QP-structure. When $N$ has boundary, we can impose appropriate {\bf boundary conditions}. One possibility comes from the choice of a lagrangian $Q$-submanifold on the target $i:\cL \hookrightarrow \M$ such that $i^*\lambda=0$ and defining
 \[ \F := \{ \phi\in Map(T[1]N, \cM)\ | \ \phi(T[1]\partial N)\subset \cL\}.\]
 We thus get that $(\F,\omega_\F,S_\F)$ defines a QP-structure (we omit the obvious pullbacks along $\F\hookrightarrow Map(T[1]N,\M)$). When we need to highlight the source and target in this construction we will write 
 \[\F\equiv \F_{N}(\M). \]
We notice that, when taking into account internal $\Z$-gradings, one conventionally considers $dim(N)=k+1$ so that the resulting QP-structure on $\F_N(\M)$ is of internal degree $-1$ as in the convention for the BV-formalism.

Finally, we observe that when $N$ has no boundary, the AKSZ action $S_\F$ is independent of the choice of potential $\lambda$. But in the presence of boundary, $\partial N\neq \emptyset$, $S_\F$ does depend on $\lambda$ and thus it has to be considered as part of the defining data for $\F$, see \cite{cat:aksz, mnev:hol}.

\bigskip

The two main cases of this paper are the Poisson sigma model and the Courant sigma model, that we describe using the AKSZ construction as follows.

\smallskip

\noindent{\bf Poisson sigma model (PSM).} 
Let $(M, \pi)$ be a Poisson manifold. We define the exact QP-mainfold $(T^*[1]M,\ \omega_{can}=d\lambda_{can},\ \theta=\pi)$ where $T^*[1]M=(M, C_\bullet^\infty(T^*[1]M)=\fX^\bullet(M)),\ \omega_{can}$ of internal degree $k=1$ (here $\lambda_{can}$ is the Liouville 1-form) and $\theta=\pi\in\fX^2(M)=C^\infty_2(T^*[1]M)$. Using coordinates $\{y^i\}$ on $M$, a coordinate description of the exact QP-manifold is
$$\fun^\infty(T^*[1]M)\supset \{y^i, v_i\} \quad \text{ with degree }\quad  |y^i|=0, \ |v_i|=1 \text{ then }$$ 
$$ \omega_{can}=dv_i dy^i,\quad \lambda_{can}=v_i d y^i,\quad Q=\pi^{ij}v_i\frac{\partial}{\partial y^j}+\frac{1}{2}\partial_i\pi^{jk}v_jv_k\frac{\partial}{\partial v_i}, \quad \theta=\frac{1}{2}\pi^{ij}v_iv_j.$$

Let $\Sigma$ be an oriented 2-dimensional manifold.   The induced space of fields for the PSM $\F\equiv \F_\Sigma(T^*[1]M)=Map(T[1]\Sigma, T^*[1]M)$ can be described by the coordinate superfields 
\begin{equation}
 \begin{array}{llc}
  ev^*y^i=\bY^i=Y^i_0+Y^i_1+Y^i_2, &|Y^i_j|=-j,& Y_0:\Sigma\to M, \quad Y_j\in\Omega^j(\Sigma; Y^*_0TM),\\
   ev^*v_i=\bV_i=
V^0_i+ V^1_i+V^2_i, &|V^j_i|=1-j, &V^j\in\Omega^j(\Sigma; Y^*_0T^*M).
 \end{array}
\end{equation}
Let us describe the components in this simple case. The fields with $|\cdot|=0$ are called classical fields, here $Y_0$ and $V_1$. Gauge symmetries are encoded by fields with degree $|\cdot|=1$, which in this case corresponds to $V^0$. The other three are called antifields, they are conjugate to the former with respect to the symplectic form 
$$ \omega_\F=\int_{T[1]\Sigma}d \bV_id \bY^i=\int_\Sigma dV_i^0 dY^i_2+ dV_i^1 dY^i_1+dV_i^2 dY^i_0.$$  
The action $S_\F$ is locally given by
\begin{equation}\label{Act-PSM}
\begin{split}
S_\F=&\int_{T[1]\Sigma} \bV_i  d_{\Sigma}\bY^i+\frac{1}{2}\pi^{ij}(\bY)\bV_i\bV_j\\
=&\int_\Sigma\Big( V^1_i  d_{\Sigma}Y^i_0+\frac{1}{2}\pi^{ij}(Y_0)V_i^1V_j^1\Big)
+ V_j^2\pi^{ij}(Y_0)V_j^0+Y^i_1\Big(d_{\Sigma}V^0_i+\partial_i\pi^{jk}(Y_0)V_j^0V_k^1\Big)\\
&+\frac{1}{2}Y^i_2\partial_i\pi^{jk}(Y_0)V_j^0V_k^0+\frac{1}{4}Y^i_1Y^j_1\partial_i\partial_j\pi^{kl}(Y_0)V_k^0V_l^0.\end{split}
\end{equation}
If $\partial\Sigma\neq\emptyset$, we choose $i:\cL\to T^*[1]M$ Lagrangian $Q$-submanifold with $i^*\lambda_{can}=0$. For example, the choice considered in \cite{cat:kon} to compute the Kontsevich $\star$-product using a disk is $\cL=0_M=\{ v^i=0\}$, or in terms of the superfields $\{\bV^i_{|T[1](\partial\Sigma)}=0\}$.

\medskip

\noindent{\bf Courant sigma model (CSM).} In Section \ref{subsec:super}, we saw that Courant algebroids $(E\to M, \langle\cdot,\cdot\rangle, \Cour{\cdot,\cdot},\rho)$ admit a description in terms of QP-manifolds $(\M,\omega,\theta)$ with $\omega$ of internal degree $2$, see Proposition \ref{sev-roy-corr}.  As mentioned in the introduction, we specialize to Courant algebroids that are the double of Lie bialgebroids. Recall that the double of a Lie bialgebroid $(A[1], Q_A, \pi_A)$ is given by the exact QP-manifold
\begin{equation}\label{iso}
(\cM=T^*[2]A[1],\ \omega=d\lambda,\ \theta=Q_A+\pi_A),
\end{equation}
with functions given by $C^\infty_k(\cM)=\oplus _{2i+j=k}\fX^i_j(A[1])$ so that $\theta\in C^\infty_3(\cM)$ (see Section \ref{subsec:super}). The Liouville 1-form $\lambda$ above is the one corresponding to the cotangent bundle fibration $T^*[2]A[1] \to A[1]$.



To fix ideas, we provide a coordinate description of \eqref{iso} as follows. The manifold $A[1]$ admits coordinates
	$$\fun^\infty(A[1])\supset \{ x^i, a^\alpha\}\quad\text{ with degree }\quad |x^i|=0,\ |a^\alpha|=1$$
	therefore the manifold $T^*[2]A[1]$ has coordinates \begin{equation}\label{coor-M}
	 \fun^\infty(T^*[2]A[1])\supset\{ x^i, a^\alpha, b_\alpha, p_i\} \quad \text{ with degree }\quad |x^i|=0,\ |a^\alpha|=1,\ |b_\alpha|=1,\ |p_i|=2;
\end{equation}	 
	the canonical symplectic form $\omega$ and the degree $3$ function are given by
\begin{equation}\label{lam-S-M}
\omega= dp_i d x^i+ db_\alpha d a^\alpha\quad\text{ and}\quad \theta=Q_A+\pi_A=\rho^i_\alpha a^\alpha p_i+\frac{1}{2} c^{\gamma}_{\alpha\beta}a^\alpha a^\beta b_\gamma+\widetilde{\rho}^{\alpha i}b_\alpha p_i+\frac{1}{2}\widetilde{c}^{\alpha\beta}_\gamma b_\alpha b_\beta a^\gamma,
\end{equation}	
with $\rho^i_\alpha, c^{\gamma}_{\alpha\beta},\widetilde{\rho}^{\alpha i},\widetilde{c}^{\alpha\beta}_\gamma \in C^\infty(M).$ Since $T^*[2]A[1]\cong T^*[2]A^*[1]$ we get that $\omega$ admit two different potentials but we just consider  $$\lambda=p_idx^i+b_\alpha da^\alpha.$$	It is easy to check that the fibre through zero 
\begin{equation}\label{boundary-M}
\quad A^*[1]\simeq \{ a^\alpha=0, p_i=0\} \hookrightarrow T^*[2]A^*[1]
\end{equation}
is a Lagrangian $Q$-submanifold satisfying $\lambda_{|A^*[1]}=0$. Therefore, it can be used as a boundary condition for the resulting topological sigma model.

Let $N$ be an oriented 3-dimensional manifold.  The induced space of superfields for the CSM is  
\begin{equation}\label{Fi-CSM}
\F\equiv \F_N(\cM)=Map(T[1]N, T^*[2]A[1])
\end{equation}
and locally it is described by the coordinate superfields\footnote{As in the PSM one can decompose the superfields into its homogeneous components.} 
$$ \bX^i=ev^*x^i, \quad \bA^\alpha=ev^*a^\alpha, \quad \bB_\alpha=ev^*b_\alpha\quad \text{and}\quad \bP_i=ev^*p_i, \text{ with }ev:T[1]N\times \F_N(\cM) \to \M.$$

The symplectic form $\omega_\F$ and the action $S_\F$ are locally given by
$$ \omega_\F=\int_{T[1]N}d \bP_i d \bX^i+d \bB_\alpha d \bA^\alpha \qquad \text{and}$$
\begin{equation}\label{Ac-CSM}
\begin{split}
S_\F=\int_{T[1]N} \Big(\bP_i&  d_N\bX^i+ \bB_\alpha d_N\bA^\alpha +\rho^i_\alpha(\bX) \bA^\alpha \bP_i+\frac{1}{2} c^{\gamma}_{\alpha\beta}(\bX)\bA^\alpha \bA^\beta \bB_\gamma\\
&+\widetilde{\rho}^{\alpha i}(\bX)\bB_\alpha \bP_i+\frac{1}{2}\widetilde{c}^{\alpha\beta}_\gamma(\bX) \bB_\alpha \bB_\beta \bA^\gamma\Big).\end{split}
\end{equation}

When $\partial N\neq\emptyset$, we choose $i:\cL\to \cM$ a Lagrangian $Q$-submanifold with $i^*\lambda=0$. In our case we will consider $\cL=A^*[1]=\{ a^\alpha=p_i=0\}$ or, in terms of the superfields, the boundary condition
$$\{ \bA^\alpha_{|T[1]\partial N}=0, \ \bP_{i|T[1]\partial N}=0\}.$$

\section{Some properties of the AKSZ construction}
In this section, we study two special properties of the AKSZ construction which are key steps in the proof of the main claim \ref{main}. Namely, the behavior under exponential type identifications (corresponding to Step (1) in eq. \eqref{eq:maincompu}) and the relation to symplectic reduction of the target space (corresponding to Step (2) in eq. \eqref{eq:maincompu}).

\subsection{The exponential map for product sources}\label{S3}

Here we discuss a general exponential-type feature of the AKSZ construction and apply it to the case of the CSM with source being $N=\Sigma \times I, \ I=[0,1]$.

First, consider a product supermanifold $\N=\N_1 \times \N_2$ and any $\M$. Then, there is a natural \emph{exponential map} identification 
\begin{equation}\label{eq:exp} Map(\N_1 \times \N_2, \M) \simeq Map(\N_1, Map(\N_2,\M)). \end{equation}
When $\N_1 \times \N_2$ is endowed with a product Berezinian volume, this exponential map interacts with the transgression in the following way
\[ \bT_{\N_1\times \N_2} \simeq \bT_{\N_1} \circ \bT_{\N_2}. \]
\begin{example}
To fix ideas, we verify the above isomorphisms in the case $\cN_1=\cN_2=\bR[1]$. Denote the coordinates on $\cM$ by $x^\alpha$ and the coordinates on $\cN_i$ by $\xi^i\in C^\infty_1(\cN_i)$. Hence, the supermanifold $Map(\N_2,\M)$ is parametrized by superfields $X^\alpha=a^\alpha+b^\alpha \xi^2$. Therefore the supermanifolds  $Map(\N_1 \times \N_2, \M)$ and $Map(\N_1, Map(\N_2,\M))$ are respectively parametrized by
$$ Y^\alpha=\phi_0^\alpha+\phi_1^\alpha \xi^1+\phi_2^\alpha \xi^2+\phi_{21}^\alpha \xi^2\xi^1\qquad Z^\alpha=X_0^\alpha+X_1^\alpha \xi^1=(a_0^\alpha+b_0^\alpha\xi^2)+(a_1^\alpha+b_1^\alpha\xi^2)\xi^1$$
inducing the desired isomorphism of \eqref{eq:exp} in an obvious way from $Y^\alpha\simeq Z^\alpha$. Now, suppose that we have $\omega=\omega_{\alpha\beta}(x)\delta x^\alpha \delta x^\beta\in\Omega^2(\cM)$. The corresponding transgression gives
\begin{equation*}
\begin{split}
\bT_{\N_1\times \N_2}(\omega)=&\int_{\cN_1\times \cN_2}\mu_{\cN_1\times\cN_2}\quad ev^*\omega=\int d\xi^1 d\xi^2\omega_{\alpha\beta}(Y)\delta Y^\alpha \delta Y^\beta\\
\simeq&\int d\xi^1\int d\xi^2\omega_{\alpha\beta}(X_0+X_1\xi^1)\delta\big(X_0^\alpha+X_1^\alpha \xi^1\big)\delta\big(X_0^\beta+X_1^\beta \xi^1\big)=\bT_{\cN_1}(\bT_{\cN_2}(\omega)).
\end{split}
\end{equation*}
\end{example}

An analogous property holds for the lifts of vector fields from the source and target to the space of maps. Thus, altogether, we have the following general property of the AKSZ construction: the exponential identification \eqref{eq:exp} induces an isomorphism
$$ \F_{N_1\times N_2}(\M) \simeq \F_{N_1}(\F_{N_2}(\M)) \text{ as (formal) QP-manifolds.} $$

\begin{remark} (Corners I)\label{rmk:corners1}
We observe that the above identification can be used to define AKSZ field theories in which the source $N$ is a manifold with corners. Concretely, when $N=N_1\times N_2$ and both the $N_j$ have boundaries, we can impose appropriate boundary conditions (as in Section \ref{gen}) following the right-hand side of the above isomorphism, as follows. First, consider a lagrangian $\cL\hookrightarrow\M$ which we use to define boundary conditions for $\partial N_2$,
$$ \F_{N_2}^{\cL}(\M):=\{\varphi \in Map(T[1]N_2,\M):\varphi|_{T[1]\partial N_2} \subset \cL \}.$$
Then, we choose a lagrangian $\L_2 \hookrightarrow \F_{N_2}^{\cL}(\M)$ which we use to fix boundary condition for $\partial N_1$: the space of fields on $N_1\times N_2$ with corner-conditions can be defined as the subset of $Map(T[1]N_1,Map(T[1]N_2,\M)) \simeq Map(T[1](N_1\times N_2),\M)$ given by
$$ \{ \phi\in Map(T[1]N_1,\F_{N_2}^{\cL}(\M)): \phi|_{\partial N_1}\subset \L_2 \}.$$
\end{remark}

\medskip

\noindent{\bf CSM on cylinders.} The following particular case is the one that we are most interested in. Consider $\M=T^*[2]A[1]$ the target space of the CSM and $N=\Sigma\times I$ as source, with $\Sigma$ a closed oriented surface and $I=[0,1]$. From the general property of the exponential map \eqref{eq:exp}, we have an identification of QP-manifolds
\[ \F_{\Sigma \times I}(\M) \simeq \F_{\Sigma}(\F_I(\M)), \]
where we choose the boundary condition corresponding to $\partial I= \{0\}\cup\{ 1\}$ by the lagrangian $\cL= A^*[1]\hookrightarrow \M$. In this way, we have
\[ \F_{\Sigma \times I}(\M) = \{\phi \in Map(T[1]\Sigma \times T[1]I,T^*[2]A[1]): \phi|_{t=0,1}(T[1]\Sigma)\subset A^*[1] \} ,\]
and
\[ \F_I(\M) = \{\gamma\in Map(T[1]I,T^*[2]A[1]): \gamma|_{t=0,1}\in  A^*[1] \}. \]
The QP-manifold $(\F_I(\M), \omega_{\F_I(\M)}, S_{\F_I(\M)})$ will play an important role in the Lie-theoretic results of the following section. 
For the reader's convenience, we summarize its relation to the CSM on the cylinder defined by $\F_{\Sigma\times I}(\M)\simeq \F_\Sigma(\F_I(\M))$ as follows: the quantum computations yield
\begin{equation*}
\langle \cO\rangle_{CSM(\Sigma\times I,A\oplus A^*)}:=\int_{\mathfrak{L}\subset \F_{\Sigma\times I}(\M)}\sqrt{\mu}\ \cO\ e^{\frac{i}{\hslash}S_{\F_{\Sigma\times I}(\M)} }=\int_{\mathfrak{L}'\subset\F_\Sigma(\F_I(\M))}\sqrt{\mu'}\ \cO'\ e^{\frac{i}{\hslash}S_{\F_\Sigma(\F_I(\M))}},
\end{equation*} 
where $\L'$, $\cO'$ and $\mu'$ denote the induced isomorphic structures under the identification \eqref{eq:exp}, and we have
\[ \omega_{\F_{\Sigma}(\F_I(\M))}=\bT_\Sigma(\omega_{\F_I(\M)}), \ S_{\F_\Sigma(\F_I(\M))}=\bT_\Sigma(S_{\F_I(M)})+i_{\widehat{d_\Sigma}}\bT_\Sigma(\bT_I(\lambda)). \]
 
\begin{remark} (Corners II)\label{rmk:corners2}
Following the general discussion of Remark \ref{rmk:corners1}, when $\partial \Sigma\neq \emptyset$, we can define $\F_{\Sigma \times I}(\M) \simeq \F_{\Sigma}(\F_I(\M))$ by taking $\F_I(\M)$ exactly as above and also choosing an appropiate lagrangian $\L_2 \hookrightarrow \F_I(\M)$ to impose the boundary conditions corresponding to $\partial \Sigma$. We will provide a concrete example of such a choice when discussing the underlying effective PSM in Section \ref{subsec:rft} below (see Remark \ref{rmk:corners3}).
\end{remark}

\subsection{Homological reduction of target spaces in AKSZ}\label{sec:redaksz}
In this subsection, we revisit from \cite{ale:red} the idea of considering AKSZ constructions in which the target space $\cM_{red}=\cM//\cC$ can be obtained as a \emph{symplectic reduction} from $\cM$ by a coisotropic $\cC\hookrightarrow \cM$. The idea is to relate such a field theory to another one constructed using as target a larger (``BFV-type") supermanifold $\cZ$ in which the reduction is encoded homologically. See also \cite{sig:tes}. 

A concise way to characterize such homologically encoded reduction is by considering an extra grading denoted by $ga(\cdot)$ (``ghost-antighost" degrees).

\begin{definition}\label{def:ga}
	Let $(\cZ, \omega, \theta)$ be a QP-manifold of internal degree $n$. We say that it carries a compatible $ga(\cdot)$ degree if it admits an atlas carrying an extra grading, called $ga(\cdot)$, in which the coordinates are $ga-$homogeneus, $ga(\omega)=0$ and $\theta=\sum_{r\leq 1} \theta_r$ with $ga(\theta_r)=r$.
\end{definition}

Indeed, following \cite{ale:red}, every such $(\cZ,\omega,\theta)$ encodes a tuple $(\cZ_0, \omega_0, \cC,\theta_0)$, which we call {\bf reduction data}, consisting of a symplectic manifold $(\cZ_0, \omega_0)$ of internal degree $n$, a coisotropic $\cC\hookrightarrow \cZ_0$ (defined by a coisotropic vanishing ideal in $C^\infty(\cZ_0)$) and a function $\theta_0\in C^\infty_{n+1}(\cZ_0)$ which is $\cC$-reducible and defines a QP-structure after reduction, as follows. 

\smallskip

Let us denote the $ga-$homogeneous coordinates on $\cZ$ by $\{q^i,\xi^a, p_{\xi^b}\}$ with $ga(q^i)=0$, $ga(\xi^a)>0$ (``ghosts") and $ga(p_{\xi^b})<0$ (``antighosts"). The coordinates with $ga= 0$ defines a submanifold \begin{equation}\label{Y0}
i:\cZ_0 \hookrightarrow \cZ\quad \text{that together with}\quad \omega_0:= i^*\omega \quad \text{and} \quad  \theta_0:= i^*\theta_{r=0}
\end{equation} 
will define part of our reduction data. The coisotropic submanifold $\cC\hookrightarrow \cZ_0$ is given locally by  
\begin{equation}\label{C}
\cC\overset{loc}{=}\{ i^*\frac{\partial \theta_1}{\partial \xi^a}=0 \text{ with } ga(\xi^a)=1\}.
\end{equation}
Although $\cC$ is only defined through a (coisotropic) vanishing ideal $I_{\cC}\subset C^{\infty}(\cZ_0)$ generated by the above functions extracted from $\theta_1$, we shall assume that $I_{\cC}$ is regular enough and that thus defines a smooth submanifold $\cC \hookrightarrow \cZ_0$.
The compatibility between $\theta_0$ and $\cC$ thus reads $\{\theta_0,I_{\cC} \}_{\cZ_0}\subset I_{\cC}$ (we say that $\theta_0$ is \emph{reducible} by $\cC$) and that $\{\theta_0,\theta_0 \}_{\cZ_0}\in I_{\cC}$.

These properties ensure that, when the symplectic quotient $\cC\to \cZ_{red}=\cZ_0//\cC$ by the singular foliation of $\omega_0|_\cC$ is regular, then $\omega_0$ and $\theta_0$ descend to $\cZ_{red}$ defining a {\bf reduced QP-manifold} $(\cZ_{red}, \omega_{red}, \theta_{red})$ (also of internal degree $n$). To see this we recall that the reduced Poisson algebra can be described as $C^\infty(\cZ_{red})=\{f\in C^\infty(\cZ_0): \{f,I_{\cC}\}\subset I_{\cC} \}/I_{\cC}$. It is interesting to notice that all the above conclusions about reduction of $\cZ_0$ come from the equation $\{\theta,\theta\}=0$ on $\cZ$ and the extra $ga$-grading.

\medskip

We say that $(\cZ,ga)$ provides an {\bf homological model} for the reduced QP-structure on $\cZ_{red}$. The paradigmatic case, which explains the nomenclature, is recalled in the next example.

\begin{example}\label{ex:standardbfv}
	The standard ``BFV" (sometimes also called ``BRST") model for (regular) reduction of an ordinary symplectic manifold $M$ by a hamiltonian $G$-action is a particular case of the above. In this case, $\cZ = M \times T^*\g[1]=M \times \g[1] \times \g^*[-1]$ where the degree shifts coincide with the $ga$-grading (so that $\cZ_0=M$) and $\g=Lie(G)$. In this case the symplectic form is the product of the form on $M$ times the canonical on $T^*\g[1]$ and
	$$\theta=\theta_{ga=1}=\mu_\alpha \xi^\alpha+\frac{1}{2}c_{\alpha\beta}^\gamma\xi^\alpha\xi^\beta p_\gamma$$ where $\mu_\alpha$ are the components of the moment map, $\xi^\alpha\in C^\infty_1(\g[1])$ and $p_\alpha\in C^\infty_{-1}(\g^*[-1])$. In particular, $\theta_0=0$ and the $ga=0$ cohomology yields the reduced Poisson algebra, $$H^{ga=0}(C^\infty(\cZ),\{ \theta,-\}_{\cZ}) = C^\infty(M_0//G).$$ See more details in \cite{ale:red} and references therein.
\end{example}

In the case of a general $(\cZ,ga)$, there is a map from certain cocycles in $(C^\infty(\cZ),\{\theta,-\}_{\cZ})$ to cocycles in the complex of the reduced QP-manifold $(C^\infty(\cZ_{red}),\{\theta_{red},-\}_{\cZ_{red}})$. Indeed, consider a cocycle of the following form
\begin{equation}\label{eq:bfvobs}
O = \sum_{r\leq 0} O_r \in C^\infty(\cZ), \ ga(O_r)=r, \ \{\theta, O \}=0.
\end{equation}
Then, $f:=O_0|_{\cZ_0} \in C^\infty(\cZ_0)$ is a \emph{reducible classical observable}, namely, $\{f,I_{\cC}\}\subset I_{\cC}$ and $\{\theta_0,f\} \subset I_{\cC}$. This implies that, $f|_{\cC}$ is basic for the quotient $\cC\to \cZ_{red}$ inducing a function $O_{red}\in C^\infty(\cZ_{red})$ which is a cocycle, $\{\theta_{red},O_{red} \}_{\cZ_{red}}=0$. In general, when $\theta_0\neq 0$, the assignment $O\to O_{red}$  might not descend to cohomology, see \cite[ Section 3, Rmk. 9]{ale:red}. On the other hand, the idea is that, under certain regularity assumptions, any such $O_{red}$ can be homologically represented by an observable $O$ in the model given by $\cZ$ thus defining a ``lifting" map from the QP-cohomology of $\cZ_{red}$ into the cohomology of $\cZ$ (see \cite[Prop. 11(ii)]{ale:red}).

\medskip

\noindent {\bf Induced AKSZ theories from homological reduction of target spaces.}
We now turn to the relation found in \cite{ale:red} between the AKSZ construction and the target space QP-reduction. Assume that the reduction $(\cZ_{red}, \omega_{red}, Q_{red}, \theta_{red})$ is regular, as above, and that $\omega=d\lambda$ with $\lambda$ reducible to $\cZ_{red}$. Hence we have two natural $(n+1)$-dimensional AKSZ sigma models with space of fields given by 
$$\F\equiv \F_N(\cZ)=Map(T[1]N, \cZ)\quad\text{and}\quad \F_{red}\equiv \F_N(\cZ_{red})=Map(T[1]N, \cZ_{red}).$$ 
The main result of \cite{ale:red} is that $\F_{red}$ arises as an effective theory coming from $\F$, as we shall review in the remaining of this section.

First, a key observation is that the $ga$ grading induces an analogous one on the space of superfields $\F$ and that the resulting structure (formally) fulfills Definition \ref{def:ga} in the case of internal degree $n=-1$. Moreover, the reduction data underlying $(\F,ga)$ yields $\F_{red}$ as the resulting reduced QP-manifold (of degree $-1$). We can thus say that $\F$ provides an homological model for the QP-structure of $\F_{red}$. 

We observe that the $ga=0$ sector underlying $\F$ is given by $\F_0:=Map(T[1]N,\cZ_0)$. Also, that (formally) we can relate \emph{classical} observables on $\F$ and $\F_{red}$ following the general homological prescription recalled above. In particular, a classical observable $\cO\in C^\infty(\F), \ \{S_{\F},\cO \}_{\F}=0$ with $ga$-structure given as in eq. \eqref{eq:bfvobs} can be seen as a representative of an observable $\cO_{red}\in C^\infty(\F_{red})$. Moreover, formally, any $\cO_{red}$ can be lifted to such a representative $\cO$ under appropriate regularity assumptions.

\subsubsection{\bf Expectation values, compatible gauge fixings and quantum observables}\label{subsub:fixandobs}
Next, following \cite[\S 4]{ale:red}, the idea is to enhance this homological model for reduction from the QP-structure to the \emph{BV-structure} (i.e. the SP-structure and its induced BV-laplacian, as recalled in Section \ref{subsec:BV}), thus allowing to relate expectation values of observables in $\F$ and $\F_{red}$. Since the involved supermanifolds are infinite dimensional, the discussion involving Berezinians and integration is formal and proceeds by analogy with finite dimensional models like the following one. 

\begin{remark} (A simple model from \cite[\S 4.1]{ale:red})\label{rmk:simpleBV}
	Let $q:N \to N/G$ be a finite dimensional principal $G$-bundle. Then $\F_0=T^*[1]N$ has a canonical degree $-1$ symplectic structure and a volume form $\lambda$ on $N$ induces a Berezinian $\mu_\lambda$ on $\F_0$ defining an SP-structure (see \cite[\S 2.1]{ale:red}). The $G$-action on $N$ naturally lifts to a hamiltonian action on $\F_0$ and the corresponding degree $-1$ symplectic reduction yields $\F_{red}=\F_0//G \simeq T^*[-1](N/G)$. This reduction can be homologically encoded in 
	$$ \F:= \F_0 \times \g[1] \times \g^*[-2]$$
	where the $ga$-grading is $1$ on the second factor and $-1$ on the third. The factor $\g[1] \times \g^*[-2]$ can be endowed with a natural Berezinian $\mu_{gh}$. Moreover, there is a natural function $\theta_{ga=1}\in C^\infty(\F)$ determined by the $G$-action (with an analogous expression to that of Example \ref{ex:standardbfv}) and, given $S_0$ a $G$-invariant degree zero function on $\F_0$ (we denote its pullback to $\F$ with the same notation), then
	$$ S := S_0 + \theta_1 \in C^\infty(\F), \ \{S,S\}_{\F}=0.$$
	This structure $(\F,ga)$ fullfills Definition \ref{def:ga} and the underlying $n=-1$ reduced QP-structure is $\F_{red}$ with the induced function $S_{red}$ naturally coming from the $G$-invariant $S_0$. This provides a simple model to analyze reduction in the context of the BV-formalism, the main new ingredient being the Berezinian $\mu_\lambda\times \mu_{gh}$ which we will study below.
\end{remark}

In order to study computations of expectation values in $\F$, we first discuss the implications of the QME to the underlying reduction. If the given action $S_\F$ (coming from the AKSZ construction with target $\cZ$) satisfies the QME, this (formally) implies that both $S_\F|_{ga=0}$ and the (formal) Berezinian $\mu$ on $\F$ descend to $\F_{red}$. This is explained in \cite[\S 4.1]{ale:red} and can be tested in the simple model of Remark \ref{rmk:simpleBV} in which one finds that 
\begin{equation}\label{eq:lambdared} i_{\delta_{\fC}} \lambda = q^* \lambda_{red},\end{equation} 
where $\delta_{\fC}$ is a fermionic Dirac delta function on $T^*[-1]N$ (i.e. a multivector on $N$) supported on the (odd) conormal $\fC\hookrightarrow T^*[-1]N$ to the $G$-orbits on $N$ and $\lambda_{red}$ is an induced volume form on $N/G$. Notice that $\fC$ is the coisotropic given as the zero level set of the underlying moment map which defines the symplectic reduction.

Similarly, we have that a quantum observable of the form $\cO=\sum_{r\leq 0} \cO_{ga=r}$  for the $\F$-theory, in particular, induces a reduced observable $\cO_{red}$ in $\F_{red}$ (recall eq. \eqref{eq:bfvobs}). More consequences of the condition of being a quantum observable in $\F$ can be found in \cite[\S 4.1]{ale:red}.

\smallskip

\noindent {\bf Compatible gauge fixings.} For the expectation value computation on $\F$ to descend to a corresponding computation on $\F_{red}$, we need to choose the gauge fixing in a particular way. The key point behind this choice is to take advantage of the $ga$-grading structure and, upon integration of ghost fields, to produce a Dirac delta supported on the coisotropic $\fC=Map(T[1]N,\cC)$.

 In the simple model of Remark \ref{rmk:simpleBV}, this choice is given by
$$ \L = N^*[-1]\Gamma \times \g[1] \times 0 \hookrightarrow T^*[-1]N \times \g[1] \times \g^*[-2]=\F$$
where $\Gamma \subset N$ is a lift of a submanifold $\widetilde{\Gamma}\subset N/G $, i.e. $q|_{\Gamma}:\Gamma\to\widetilde{\Gamma}$ is a (local) diffeomorphism, and the coordinates $p_a\in C^\infty(\g^*[-2])$ with $ga=-1$ (i.e. the antighosts) are set to zero. With this choice, for $\cO\in C^\infty(\F)$ as in eq. \eqref{eq:bfvobs} and satisfying the quantum observable condition $\Delta(\cO e^{\frac{i}{\hbar}S})=0$, the expectation value computation yields (see \cite[Prop. 15]{ale:red})
\begin{equation}\label{obs-fd}
\begin{split}
\langle \cO \rangle = &\int_{\L}\sqrt{\mu_\lambda \times \mu_{gh}}\  \cO\ e^{\frac{i}{\hbar}S}  = \int_{ N^*[-1]\Gamma} \sqrt{\mu_\lambda} \ \delta_{\fC}\ \cO_0|_{\F_0} \ e^{\frac{i}{\hbar}S_0} \\ 
= &\int_{N^*[-1]\widetilde{\Gamma}}\sqrt{\mu_{\lambda_{red}}}\ \cO_{red}\ e^{\frac{i}{\hbar}S_{red}} = \langle \cO_{red}\rangle,
\end{split} 
\end{equation}
where, in the second equality, we integrated out the ghost $ga=1$ variables $\xi^a$ (corresponding to $\g[1]$) producing the (fermionic) delta function $\delta_{\fC}$ supported on $\fC\hookrightarrow T^*[-1]N$ out of the term $\theta_1|_{p_a=0}=\mu_a \xi^a$ in the exponent (see also Example \ref{ex:standardbfv}), and we then used equation \eqref{eq:lambdared} relating the volumes $\lambda$ and $\lambda_{red}$ on $N$ and $N/G$, respectivelly. It is interesting to notice that, $\Gamma$ being a cover for $\widetilde{\Gamma}$, entails an underlying transversality condition on the $ga=0$ part of the gauge fixing, $\L_0:=N^*[-1]\Gamma$, with respect to the quotient map $\cC\to T^*[-1](N/G)$. This condition ensures that $\L_0$ descends to the well defined lagrangian $\L_{red}:=N^*[-1]\widetilde{\Gamma}\hookrightarrow \F_{red}$ and is essential for $\langle \cO \rangle$ not to be artificially zero (or ill defined in an infinite dimensional situation).

Generalizing the situation given in the model above, coming back to $\F=Map(T[1]N,\cZ)$, the choice of \emph{compatible gauge fixing} $\L\hookrightarrow \F$ is taken as follows: the $ga<0$ fields are set to zero (i.e. $ev^*p_{\xi^b}=0$) and $\L_0:=\L\cap \F_{ga=0}$ (namely, the $ga=0$ sector of the gauge fixing) is required to descend to a lagrangian $\L_{red}\hookrightarrow \F_{red}$ under reduction. Thus, the lagrangian $\L$ will locally look like $\L_0 \times \L_{gh}$ with $\L_0 \subset Map(T[1]N,\cC)$ satisfying a transversality condition relative to the quotient $\cC\to \cZ_{red}$ and where $\L_{gh}$ is defined by $ev^*p_{\xi^b}=0$ as above.

\smallskip

\noindent{\bf Expectation values.} With a compatible gauge fixing $\L\hookrightarrow \F$ as above and considering a quantum observable $\cO \in C^\infty(\F)$ with the $ga$-structure as in \eqref{eq:bfvobs}, we formally obtain
	\begin{equation*}
	\begin{split}
	\langle \cO\rangle=&\int_{\mathfrak{L}\subset \F}\sqrt{\mu}\ \cO\ e^{\frac{i}{\hslash}S_\F}\overset{(*)}{=} \int_{\L_0\subset \F_0}\sqrt{\mu_0}\ \delta_{\frak{C}} \ \cO|_{ga=0}\ e^{\frac{i}{\hslash}S_{\F}|_{ga=0}}\\
	 =&\int_{\mathfrak{L}_{red}\subset\F_{red}}\sqrt{\mu_{red}}\ \cO_{red}\ e^{\frac{i}{\hslash}S_{\F_{red}}}=\langle \cO_{red}\rangle
	\end{split}
	\end{equation*} 
The step $(*)$ above is a pushforward in which the $ga>0$ superfields corresponding to $ev^*\xi^a$ are integrated out and the result is a delta distribution factor $\delta_{\mathfrak{C}}$ supported on $\mathfrak{C}=Map(T[1]N,\cC)$ (in general, it can have both bosonic and fermionic components). By the reducibility of $S_\F|_{ga=0}$ and $\cO|_{ga=0}$ (coming from the QME and the quantum observable condition, respectively), the resulting integral can be identified with one in associated to the symplectic quotient $\F_{red}=Map(T[1]N,\cZ_{red})$. See more details in \cite[\S 4.2]{ale:red}.

In the simple finite dimensional model recalled above, the equality of expectation values \eqref{obs-fd} is verified exactly. In infinite dimensional situations, a priori certain ``anomalies" could occur so that the computations on $\F$ and $\F_{red}$ yield different results. Nevertheless, in concrete cases, we expect that appropriate choices of the underlying regularization (of infinite dimensional integrals), of the ($ga=0$) gauge fixing, and of the homological representative $\cO$ of $\cO_{red}$, can be made so that, indeed, the independent computations of $\langle \cO\rangle$ and $\langle \cO_{red}\rangle$ coincide. 

\begin{remark}\label{rmk:cotangent}
	If one starts with the reduced data $\mu_{red}$, $\L_{red}$ and $\cO_{red}$ it can be a non-trivial task to find the corresponding $\mu$, $\L$ and $\cO$ satisfying the above equality. In particular, $\L_{red}$ has to be lifted along the symplectic reduction $\F_0 \mapsto \F_{red}$ and finding $\cO$ can be phrased in homological terms where the technique of homological perturbation is typically used (see \cite[\S 3 and \S 4]{ale:red} for a detailed discussion). 
\end{remark}

\section{Lie-theoretic result and Field-theoretic application}\label{sec-red}
In this section, we show the main Lie-theoretic result which characterizes the symplectic reduction of the target space appearing in Step (3) in eq. \eqref{eq:maincompu}. After this result is established, we finish the proof of the main Claim \ref{main} by combining all the arguments described thus far.

\subsection{Lie-theoretic interpretation of the coisotropic reduction}\label{subsec:rlie}
Let $(A[1], Q_A, \pi_A)$ encode a Lie bialgebroid $(A,A^*)$. As explained in Sections \ref{S.bia} and \ref{subsec:super}, the double Courant algebroid $E=A\oplus A^*$ corresponds to the  exact QP-manifold of internal degree $2$ given by $$\cM\equiv(\cM=T^*[2]A[1],\ \omega=d\lambda,\ \theta=Q_A+\pi_A).$$
Here, we show that $(\cZ=\F_I(\cM), \omega_\cZ, S_\cZ)$ with boundary conditions induced by $A^*[1]$ can be endowed with the structure encoding an underlying symplectic reduction (in the sense of Section \ref{sec:redaksz}) and that the associated reduced space is given by $(T^*[1]G_A, \omega_{can}, \pi_G)$, namely, precisely by the target space of a PSM on the Poisson groupoid $(G_A,\pi_G)$ integrating the lie bialgebroid $(A,A^*)$. The key argument is completely rigorous and only involves path spaces, extending the construction of $G_A$ as $A$-paths modulo $A$-homotopies of \cite{crfer} (see Section \ref{subsec:lie}). 

\medskip


\noindent{\bf Auxiliary finite-dimensional structures.} First notice that the graded manifold $T[1]I$ decompose into the product $$T[1]I=I\times \bR[1].$$ Therefore we can use the exponential identification with respect to this product. The advantage is that the graded manifold $$\cY:=\F_{\bR[1]}(\cM)=Map(\bR[1], T^*[2]A[1])=T[-1]T^*[2]A[1]$$ is finite dimensional, the only warning is that it has coordinates in positive and negative degrees. This finite dimensional $\bZ$-graded supermanifold carries the relevant $ga$-grading structure that we will need to describe $\cZ=\F_I(\cM)$ below.

We start by identifying the relevant geometric structures of the manifold $\cY$. First notice that it fits into a double vector bundle given by
\begin{equation*} 
	\xymatrix{\cY\ar[d]\ar[r]& T[-1]A^*[1]\ar[d]\\ \cM\ar[r]& A^*[1]}
\end{equation*}
with $\cM=T^*[2]A[1]$ as in \eqref{iso}.
Therefore $\cY$ can be endowed with an additional grading that we will denote by $ga(\cdot)$ coming from the above double vector bundle structure (see \cite{mac:book} for the general theory of double vector bundles). This grading is given by $-1$ for the vertical fibers and $1$ for the horizontal fibers. (In the language of double vector bundles, we thus obtain that the \emph{core} variables have $ga=0$ degree.) Notice that it is globally defined. Second, since $\bR[1]$ has a natural Berezinian we can make transgression from $\cM$ to $\cY$ of $\omega$, $\lambda$ and $\theta$ to obtain a QP-structure of degree $1$ defined by
$$\Big(\cY,\quad \omega_\cY:=\bT_{\bR[1]}(\omega)=\omega^T,\quad \lambda_\cY:=\bT_{\bR[1]}(\lambda)=\lambda^T,\quad \theta_\cY:=\bT_{\bR[1]}(\theta)=\theta^T\Big).$$ 
It is worth noting that the above transgressions coincide with an odd ($[-1]$-shifted) version of the tangent lift of differential forms $\Omega(\cM) \to \Omega(T[-1]\cM), \ \beta \mapsto \beta^T$, see \cite[\S 1.3]{yan:tan}.

\begin{remark}
A coordinate description of $(\cY,\ \omega_\cY=d\lambda_\cY,\ \theta_\cY)$ and the $ga(\cdot)$ grading is the following. The manifold $\cY=T[-1]T^*[2]A[1]$ it has coordinates $\{ q, \dot{q}\}$ for $q\in\{ x^i, a^\alpha, b_\alpha, p_i\}$ coordinates on $T^*[2]A[1]$ with 
$|\dot{q}|=|q|-1$ and  \begin{equation}\label{ga-deg}
ga(x^i)=ga(\dot{a}^\alpha)=ga(b_\alpha)=ga(\dot{p}_i)=0, \ ga(a^\alpha)=ga(p_i)=1, \ ga(\dot{x}^i)=ga(\dot{b}_\alpha)=-1.
\end{equation} 
The $1$-form $\lambda_\cY\in \Omega^1_1(\cY)$ is given by
$$ \lambda_\cY= \dot{p}_i d x^i-p_i d\dot{x}^i+ \dot{a}^\alpha d b_\alpha+ a^\alpha d \dot{b}_\alpha$$
and the function $\theta_\cY\in C^\infty_2(\cY)$ has the form $\theta_\cY=\theta^0_\cY+\theta^1_\cY$ with
\begin{equation*}
\begin{split}
\theta^1_\cY=&\partial_j\rho^i_\alpha \dot{x}^j a^\alpha p_i+\rho^i_\alpha \dot{a}^\alpha p_i-\rho^i_\alpha a^\alpha \dot{p}_i+\frac{1}{2} \partial_ic^{\gamma}_{\alpha\beta}\dot{x}^ia^\alpha a^\beta b_\gamma+c^{\gamma}_{\alpha\beta}\dot{a}^\alpha a^\beta b_\gamma+\frac{1}{2}c^{\gamma}_{\alpha\beta}a^\alpha a^\beta \dot{b}_\gamma\\
\theta^0_\cY=&\partial_j\widetilde{\rho}^{\alpha i}\dot{x}^j b_\alpha p_i+\widetilde{\rho}^{\alpha i}\dot{b}_\alpha p_i-\widetilde{\rho}^{\alpha i}b_\alpha \dot{p}_i+\frac{1}{2}\partial_i\widetilde{c}^{\alpha\beta}_\gamma \dot{x}^i b_\alpha b_\beta a^\gamma+\widetilde{c}^{\alpha\beta}_\gamma \dot{b}_\alpha b_\beta a^\gamma+\frac{1}{2}\widetilde{c}^{\alpha\beta}_\gamma b_\alpha b_\beta \dot{a}^\gamma.
\end{split}
\end{equation*}
Notice that, in the setting of Section \ref{sec:redaksz}, the ghosts $\xi^a$ correspond to $\{a^\alpha,p_i\}$ while the antighosts $p_{\xi^b}$ correspond to $\{\dot{x}^i, \dot{b}_\alpha \}$.
\end{remark}

It is easy to see that 
\begin{equation*}
ga(\omega_\cY)=0,\quad ga(\theta^1_\cY)=1 \quad \text{and}\quad ga(\theta^0_\cY)=0.
\end{equation*}
We can then extract from $\cY$ and its $ga(\cdot)$-grading the underlying reduction data as recalled in Section \ref{sec:redaksz}.

\begin{proposition}\label{Red-Y}
The reduction data $(\cY_0,\ i^*\omega_\cY,\ i^*\theta^0_\cY,\ \cC)$ is given by 
$$\cY_0=T^*[1]A, \quad i^*\omega_\cY=\omega_{can},\quad i^*\theta^0_\cY=\pi_A, \quad \cC\overset{loc}{=}\{ \rho^i_\alpha\dot{a}^\alpha=0, \ -\rho^i_\alpha \dot{p}_i+c^\gamma_{\beta\alpha}\dot{a}^\beta b_\gamma=0\}.$$
\end{proposition}
\begin{proof}
 By \eqref{Y0} we get that $\cY_0$ is locally given by the coordinates with $ga(\cdot)$ grading $0$ and in our case those are $\{ x^i, \dot{a}^\alpha, b_\alpha, \dot{p}_i\}$, see \eqref{ga-deg}. By the properties of the tangent lift, $\{x^i, \dot{a}^\alpha\}$ are coordinates on $A$ (of degree 0) and $\{ \dot{p}_i, b_\alpha\}$ are cotangent coordinates (of degree $1$), thus   $\cY_0=T^*[1]A$. This identifies $C^\infty_2(\cY_0)=\fX^2(A)$ and $\Omega^2_1(\cY_0)=\Omega^2_{lin}(T^*A)$. Working in coordinates we see that
 $$i^*\omega_\cY=d\dot{p}_idx^i+db_\alpha d\dot{a}^\alpha=\omega_{can}\quad \text{and}\quad i^*\theta_0=-\widetilde{\rho}^{\alpha i}b_\alpha \dot{p}_i+\frac{1}{2}\widetilde{c}^{\alpha\beta}_\gamma b_\alpha b_\beta \dot{a}^\gamma=\pi_A.$$
Finally, the coisotropic submanifold is given by $\eqref{C}$ and a direct computation show that it is defined by the equations $\cC=\{ \rho^i_\alpha\dot{a}^\alpha=0, \ -\rho^i_\alpha \dot{p}_i+c^\gamma_{\beta\alpha}\dot{a}^\beta b_\gamma=0\}.$
\end{proof}

One observes that the coisotropic submanifold corresponds to infinitesimal Lie algebroid paths. In this case, the reduced space $T^*[1]A//\cC$ is singular. Below, we will see that our space of interest $\F_I(\cM)\simeq Map(I,\cY)$ involves finite algebroid paths and homotopies and that, in this case, the corresponding reduction is regular (whenever $A$ is integrable).


\medskip

\noindent{\bf The reduction underlying $\F_I(\cM)$.} We reproduce the preceding computations in $$\cZ:=\F_I(\cM)=\{\Phi\in Map(T[1]I,\cM)\ |\ \Phi(T[1]\partial I)=A^*[1]\subseteq \cM\}.$$ 
Since $\cM$ is an exact QP-manifold of degree $2$ and $T[1]I$ has a $1$-Berezinian we can transgress $\omega$, $\lambda$ and $S$ and obtain a degree $1$ QP-structure on $\cZ$ defined by 
\begin{equation*}
\omega_\cZ:=\bT_I(\omega),\quad \lambda_\cZ:=\bT_I(\lambda)\quad\text{and}\quad \theta_\cZ:=i_{\widehat{d_I}}\bT_{I}(\lambda)+\bT_{I}(\theta).
\end{equation*}

The exponential property of the mapping space allows us to identify $\cZ$ with paths in $\cY$,
\begin{equation}\label{iso Y-Z}
\cZ\cong\{ \Phi\in Map(I,\cY) \ | \ \Phi_{t=0,1}\in A^*[1]\}.
\end{equation}
Moreover, the key point is that the $ga(\cdot)$ grading of $\cY$ can be transported into $\cZ$.

\begin{remark} \label{rmk:coordZ}
The local coordinates on $\cM$ given by $\eqref{coor-M}$ induce superfields on $\cZ$ of the form
\begin{equation*}
\arraycolsep=1.7pt\def\arraystretch{1.3}
\begin{array}{ccccrclrcl}
ev^*_{T[1]I}x^i&=&X^i+\dot{X}^i &\quad \text{with}\quad& X&:&I\to M, & \dot{X}&\in&\Omega^1(I; X^*TM),\\
 ev^*_{T[1]I}a^\alpha&=&A^\alpha+\dot{A}^\alpha &\text{with}&  A&\in&\Omega^0(I,X^*A),& \dot{A}&\in&\Omega^1(I; X^*A),\\
 ev^*_{T[1]I}b_\alpha&=&B_\alpha+\dot{B}_\alpha&\text{with}& B&\in&\Omega^0(I; X^*A^*),& \dot{B}&\in&\Omega^1(I; X^*A^*),\\
 ev^*_{T[1]I}p_i&=&P_i+\dot{P}_i &\text{with}& P&\in&\Omega^0(I; X^*T^*M),& \dot{P}&\in&\Omega^1(I; X^*T^*M).\\
\end{array}
\end{equation*}
Moreover, recalling the set of coordinates $\{x^i,  a^\alpha, b_\alpha, p_i, \dot{x}^i,  \dot{a}^\alpha, \dot{b}_\alpha, \dot{p}_i\}$  for $\cY$, the isomorphism \eqref{iso Y-Z} allows us to identify $X^i=ev_I^*x^i,\ \dot{X}^i=ev_I^*\dot{x}^i$ and so on. Therefore, the $ga(\cdot)$-degree of the fields are
\begin{equation*}\label{ga-deg-Z}
ga(X^i)=ga(\dot{A}^\alpha)=ga(B_\alpha)=ga(\dot{P}_i)=0, \ ga(A^\alpha)=ga(P_i)=1, \ ga(\dot{X}^i)=ga(\dot{B}_\alpha)=-1.
\end{equation*} 
The boundary conditions read
\begin{equation}\label{boun-Z}
\{ A^\alpha_{|t=0,1}=0,\quad P_{i|t=0,1}=0\},
\end{equation}
and $\omega_\cZ,$ $\lambda_\cZ$ and $\theta_\cZ=\theta_\cZ^0+\theta_\cZ^1$ are given by
\begin{equation}\label{obj-cor-Z}
\begin{split}
\omega_\cZ=&\int_I d \dot{P}_idX^i+dP_id\dot{X}^i+d\dot{B}_\alpha dA^\alpha+dB_\alpha d\dot{A}^\alpha,\\
\lambda_\cZ=&\int_I \dot{P}_idX^i-P_id\dot{X}^i+\dot{B}_\alpha dA^\alpha+B_\alpha d\dot{A}^\alpha.\\
 \theta_\cZ^1=&
 \int_I P_id_I X^i+A^\alpha (d_IB_\alpha)+ \partial_i\rho^j_\alpha \dot{X}^i A^\alpha P_j+ \rho^i_\alpha \dot{A}^\alpha P_i-\rho^i_\alpha A^\alpha \dot{P}_i\\
 &+\frac{1}{2} \partial_ic^\gamma_{\alpha\beta} \dot{X}^i A^\alpha A^\beta B_\gamma +c^\gamma_{\alpha\beta} \dot{A}^\alpha A^\beta B_\gamma +\frac{1}{2}c^\gamma_{\alpha\beta} A^\alpha A^\beta \dot{B}_\gamma 
            \\
 \theta_\cZ^0=&\int_I   \partial_i\widetilde{\rho}^{\alpha j} \dot{X}^i B_\alpha P_j+ \widetilde{\rho}^{\alpha i} \dot{B}_\alpha P_i  -\widetilde{\rho}^{\alpha i} B_\alpha \dot{P}_i\\
 &+ \frac{1}{2} \partial_i\widetilde{c}_\gamma^{\alpha\beta} \dot{X}^i B^\alpha B^\beta A_\gamma +\widetilde{c}_\gamma^{\alpha\beta} \dot{B}^\alpha B^\beta A_\gamma +\frac{1}{2}\widetilde{c}_\gamma^{\alpha\beta} B^\alpha B^\beta \dot{A}_\gamma.       
\end{split}
\end{equation}
\end{remark}

All the constructions relative to $\cZ$ have been described geometrically and we can thus reduce the verification of non-trivial properties to coordinate computations. For example, by inspection on the above formulas we get that
\begin{equation*}
ga(\omega_\cZ)=0,\quad ga(\theta^1_\cZ)=1 \quad \text{and}\quad ga(\theta^0_\cZ)=0.
\end{equation*}
As before, this structure on $\cZ$ together with the extra $ga$-grading determines symplectic reduction data as in Section \ref{sec:redaksz}.

\begin{proposition}\label{prop:Zcoisotr}
The reduction data $(\cZ_0,\ i^*\omega_\cZ,\ i^*\theta^0_\cZ,\ \cC)$ with $i:\cZ_0\hookrightarrow\cZ$ is given by 
$$\cZ_0=Map(I, T^*[1]A) ,\quad i^*\omega_\cZ=\int_I ev^*\omega_{can},\quad i^*\theta^0_\cZ=\int_I ev^*\pi_A$$
$$\cC\overset{loc}{=}\{ \partial_tX^i+\rho^i_\alpha\dot{A}^\alpha=0, \ \partial_t B_\alpha-\rho^i_\alpha \dot{P}_i+c^\gamma_{\beta\alpha}\dot{A}^\beta B_\gamma=0\}.$$
Moreover, we have that $i^*\lambda_\cZ=\int_I ev^*\lambda_{can}$ and this 1-form is reducible by $\cC$.
\end{proposition}
\begin{proof}
From the definition of the $ga(\cdot)$-degree on $\cZ$ and Proposition \ref{Red-Y} follows immediately that $\cZ_0=Map(I,T^*[1]A)$.  Using the expressions \eqref{obj-cor-Z} we obtain the other statements by straightforward computations.
\end{proof}

This proposition establishes a key connection between the structure on $\cZ$ and the Lie-theoretic integration of $A$ described in Section \ref{subsec:lie}. The first equation in the definition of $\cC$ above is no other than the equation \eqref{path-eq} for an algebroid path $TI\to A$. The second equation is the odd ($[1]$-shifted) analogue of the equation for an algebroid path in the cotangent lifted algebroid $T^*A$ (see eq. \eqref{eq:cotagpath} in Example \ref{ex:cotangpaths}). Since the cotangent lifted algebroid structure on $T^*A\to A^*$ is ``linear" (indeed, it defines a so-called VB-algebroid \cite{mac:book}), the $[1]$-shift into $T^*[1]A\to A^*[1]$ is inessential for the present considerations, and we analogously have that 
$$ Map(I, T^*[1]A)//\cC = \underset{\sim}{\underline{\{\text{algebroid paths in } T^*[1]A \}}} \simeq T^*[1]G_A.$$
Above, we use $G_A=PA/\sim$ as recalled in Section \ref{subsec:lie}. Under this identification, the function $i^*S^0_\cZ|_{\cC}$ can be identified with a lift of $\pi_A$ to the space of algebroid paths $PA$, as appears in \cite[\S 3.3]{xu:uni}, where it is also shown that it descends to $\pi_G$ (seen here as a function on $T^*[1]G_A$). In conclusion, we obtain the main result of this subsection:

\begin{theorem}\label{prop-red} Let $(A[1],Q_A, \pi_A)$ define a Lie bialgebroid $(A,A^*)$ which integrates to the Poisson groupoid $(G_A\rightrightarrows M, \pi_G)$. Then, the quotient $\cC\to \cZ_{red}$ corresponding to the reduction data of Proposition \ref{prop:Zcoisotr} is regular and the reduced QP-manifold $\big(\cZ_{red},\ (\omega_\cZ)_{red}=d(\lambda_\cZ)_{red},\ (\theta_\cZ)_{red}\big)$ of internal degree $1$ is given by
$$\cZ_{red}=T^*[1]G_A, \quad (\omega_\cZ)_{red}=\omega_{can},\quad (\lambda_\cZ)_{red}=\lambda_{can},\quad (\theta_\cZ)_{red}=\pi_G.$$
\end{theorem}

\begin{remark}\label{rmk:changeAAast}
The roles of $A$ and $A^*$ can be easily reversed in all the constructions above. In this case, the boundary condition is given by $A[1]\subseteq T^*[2]A[1]$, the zero section of the other vector bundle structure. We obtain a different $ga(\cdot)$-grading given by  
\begin{equation*}
\widetilde{ga}(X^i)=\widetilde{ga}(\dot{B}_\alpha)=\widetilde{ga}(A^\alpha)=\widetilde{ga}(\dot{P}_i)=0, \ \widetilde{ga}(B_\alpha)=\widetilde{ga}(P_i)=1, \ \widetilde{ga}(\dot{X}^i)=\widetilde{ga}(\dot{A}^\alpha)=-1.
\end{equation*} 
In order to obtain as reduced manifold $T^*[1]G_{A^*}$ with the canonical $1$-form as potential, we also need to change the potential in the Courant algebroid to the one corresponding to the cotangent fibration $T^*[2]A[1]\simeq T^*[2]A^*[1]\to A^*[1]$. In coordinates,
$$\widetilde{\lambda}=p_idx^i+a^\alpha db_\alpha.$$
\end{remark}



\begin{remark}[Lie quasi-bialgebroids]
Here we comment on the possibility of applying this sub-section's results to more general Courant algebroids than doubles of Lie bialgebroids. Pick a vector bundle $A\to M$  and consider the internal degree $2$ symplectic manifold $(T^*[2]A[1], \omega)$. From Section \ref{subsec:super} we have that a $\theta\in C^\infty_3(T^*[2]A[1])$ decomposes into $\theta=H+Q_A+\pi_A+\widetilde{H}$ with $H\in C^\infty_3(A[1])=\Gamma \wedge^3 A^*$ and $\widetilde{H}\in\fX^3_{-3}(A[1])=\Gamma \wedge^3 A$. When $\{\theta, \theta\}=0$ we have a so-called {\bf proto-Lie bialgebroid} structure on $A$,  see \cite{ive:twi, roy:qua}. Following the notation of \eqref{coor-M} the coordinate expression of the new terms are
$$ H=\frac{1}{3}h_{\alpha\beta\gamma}a^\alpha a^\beta a^\gamma\qquad\text{and}\qquad \widetilde{H}=\frac{1}{3}\widetilde{h}^{\alpha\beta\gamma}b_\alpha b_\beta b_\gamma.$$
If $H\neq 0$ then $A\to M$ is not a Lie algebroid and can not be integrated. Moreover, considering $\cZ\subset Map(T[1]I, T^*[2]A[1])$ as in this sub-section, we observe that the induced function $\theta_\cZ$ as in \eqref{obj-cor-Z} contains the extra $ga$-degree 2 term
$$\theta_\cZ^2= \int_I  \partial_ih_{\alpha\beta\gamma} \dot{X}^i A^\alpha A^\beta A^\gamma+h_{\alpha\beta\gamma} \dot{A}^\alpha A^\beta A^\gamma\quad\text{with}\quad ga(\theta^2_\cZ)=2.$$
Therefore the reduction procedure of Section \ref{sec:redaksz} does not apply to $\cZ$ since Definition \ref{def:ga} is not fulfilled.
Next, assuming $H=0$, the structure on $A$ is known as a {\bf Lie quasi-bialgebroid}. In particular,  $A$ is a Lie algebroid but $\pi_A$ is not Poisson (only quasi-Poisson) and the defect to the Jacobi identity is controlled by $\widetilde{H}$. It was shown in \cite[Thm.4.9]{xu:uni} that Lie quasi-bialgebroids integrate to {\bf quasi-Poisson groupoids}, i.e. a Lie groupoid $G$ endowed with a bivector which is \emph{not Poisson}. We thus see that the effective theory should deviate from the ordinary PSM on $G$ (the CME associated to this non-integrable bivector is not satisfied). From the point of view of our Lie-theoretic results above, in this case, we see that the Lagrangian $A^*[1]=\{a^\alpha=0, p_i=0\}\subset T^*[2]A[1]$ is no longer a Q-submanifold due to the term $\{\widetilde{H},-\}$ in the Q-structure. It is thus no longer a valid boundary condition for the definition of $\cZ$ and thus our procedure cannot be carried out (as expected since, otherwise, it would yield a solution to the CME). To deal with quasi-bialgebroids we think that a small modification of our method is needed in which one allows for boundary contributions that can eventually produce an effective theory on the quasi-Poisson groupoid similar to the twisted Poisson sigma model studied in \cite{kli:twi}. We plan to explore this elsewhere.
\end{remark}

\subsection{Field-theoretic application: the PSM on $G_A$ from a CSM on $A\oplus A^*$}\label{subsec:rft}

Here we give the details on the heuristic manipulations with path integrals involved in the main claim \ref{main} stating that the Poisson sigma model on the groupoid is an effective theory for the Courant sigma model on $A\oplus A^*$. A key point is that all the needed constructions, as explained in the previous sections, are of geometric nature and thus the relevant verifications can be reduced to the local coordinate case. Also, for the computation of expectation values we will follow the general procedure of Section \ref{subsub:fixandobs} which takes into account the $ga$-grading on $\cZ=Map(T[1]I,\M)$.

\smallskip

We begin considering our Courant sigma model as in Section \ref{gen}. Recall that for a $3$-dimensional manifold $N$, the expectation value of an observable $\cO\in C^\infty(\F)$ in the Courant sigma model $(\cM=T^*[2]A[1],\omega=d\lambda, \theta=Q_A+\pi_A)$ is given by 
$$\langle\cO\rangle_{CSM(N,A\oplus A^*)}=\int_{\L\subset \F} \sqrt{\mu}\ \cO\ e^{\frac{i}{\hslash}S_\F}$$
where the space of fields $\F$ is defined in \eqref{Fi-CSM} and the action $S_\F$ is given by \eqref{Ac-CSM}. When $N$ is the cylinder $N=\Sigma\times I$, the exponential map explained in Section \ref{S3}, yields the isomorphism
$$\F= Map(T[1](\Sigma \times I), \cM)\simeq Map(T[1]\Sigma,Map(T[1]I,\cM))=Map(T[1]\Sigma, \cZ),$$
where $\cZ=Map(T[1]I, \cM)$. Let us fix coordinates on $\cM$ as in Section \ref{gen}. The above isomorphism induces a decomposition of superfields as
$$\bX^i=\nX^i+\dot{\nX}^i,\quad \bA^\alpha=\nA^\alpha+\dot{\nA}^\alpha,\quad \bB_\alpha=\nB_\alpha+\dot{\nB}_\alpha, \quad \bP_i=\nP_i+\dot{\nP}_i$$
where $\nX^i=ev_\Sigma^*X^i$ for $X^i$ coordinate in $\cZ$ and so on. For simplicity, we assume that $\partial\Sigma=\emptyset$ 
so that $\partial N=\Sigma\times \partial I$, and we impose the boundary conditions coming  from \eqref{boun-Z}, i.e. 
$$\{ \nA^\alpha_{|t=0,1}=0,\quad \nP_{i|t=0,1}=0\}.$$

\begin{remark} (Corners III) \label{rmk:corners3}
Following Remarks \ref{rmk:corners1} and \ref{rmk:corners2}, the discussion can be easily adapted to the case $\partial \Sigma\neq \emptyset$ by choosing a lagrangian $\L_2\hookrightarrow \cZ=\F_I(\M)$ to define the boundary conditions corresponding to $\partial \Sigma$ in $Map(T[1]\Sigma, \cZ)$ on top of those corresponding to $\partial I$ already included in the definition of $\cZ$. To see a concrete example, let us recall that, when $\Sigma = D$ is a 2-disk, the boundary conditions on $T[1]\partial D$ used by Cattaneo-Felder in \cite{cat:kon} for a PSM with target a general Poisson manifold $(P,\pi_P)$ correspond to the zero section $0_P\hookrightarrow T^*[1]P$. Then, a choice of boundary conditions for our CSM on $D\times I$ which will lead to the Cattaneo-Felder BCs on the effective theory with source $D$ and target $P=G_A$ corresponds to the lagrangian $\L_2\hookrightarrow \cZ$ which, in the coordinates for $\cZ$ given in Remark \ref{rmk:coordZ}, is given by
$$ \L_2 \overset{loc}{=}\{B_\alpha =0, \dot P_i = 0, \dot X^i =0, \dot B_\alpha  =0  \}.$$
Notice that the choice $\dot X^i =0, \dot B_\alpha  =0$ corresponds to the $ga=-1$ sector and is taken for consistency with the overall gauge fixing described below. The equations $B_\alpha =0, \dot P_i = 0$ define a lagrangian on $\cZ_{ga=0}$ which goes down to the desired zero section of $T^*[1]G_A$ upon reduction.
\end{remark}

Let us next study the structure on $Map(T[1]\Sigma, \cZ)$. In the new superfields, the symplectic form is given by $$\omega_\F=\int_{T[1]\Sigma}\mu_\Sigma\int_I d\dot{\nP}_i d\nX^i+d\nP_i d\dot{\nX}^i+d \dot{\nB}_\alpha d\nA^\alpha+d\nB_\alpha d\dot{\nA}^\alpha$$ and  the action $S_\F=i_{\widehat{d_{\Sigma\times I}}}\bT(\lambda)+ \bT_{\Sigma\times I}(\theta)$ takes the form
    \begin{equation*}
         \begin{split}
             i_{\widehat{d_{\Sigma\times I}}}\bT(\lambda)=&\int_{T[1]\Sigma}\mu_\Sigma\int_{T[1]I} \mu_I\Big( \bP_i (d_\Sigma \times d_I)\bX^i+ \bB_\alpha (d_\Sigma \times d_I) \bA^\alpha\Big)\\
             =&\int_{T[1]\Sigma}\mu_\Sigma\int_I \dot{\nP}_id_\Sigma \nX^i+\nP_i d_I\nX^i-\nP_id_\Sigma \dot{\nX}^i+ \dot{\nB}_\alpha d_\Sigma \nA^\alpha-\nB_\alpha d_I\nA^\alpha+ \nB_\alpha d_\Sigma \dot{\nA}^\alpha\\
             =&\int_{T[1]\Sigma}\mu_\Sigma\int_I \Big(\dot{\nP}_id_\Sigma \nX^i+ \nB_\alpha d_\Sigma \dot{\nA}^\alpha\Big)+\nP_i d_I\nX^i+\nA^\alpha(d_I\nB_\alpha) -\nP_id_\Sigma \dot{\nX}^i+ \dot{\nB}_\alpha d_\Sigma \nA^\alpha ,
\end{split}
\end{equation*}
\begin{equation*}
\begin{split}
             \bT_{\Sigma\times I}(\theta)=&\int_{T[1]\Sigma}\mu_\Sigma\int_{T[1]I}\mu_I\Big(\rho^i_\alpha \bA^\alpha \bP_i+ \frac{1}{2} c^\gamma_{\alpha\beta} \bA^\alpha \bA^\beta \bB_\gamma +\widetilde{\rho}^{\alpha i} \bB_\alpha \bP_i+ \frac{1}{2} \widetilde{c}_\gamma^{\alpha\beta} \bB_\alpha \bB_\beta \bA_\gamma\Big)\\
             =&\int_{T[1]\Sigma}\mu_\Sigma\int_I\partial_i\rho^j_\alpha \dot{\nX}^i \nA^\alpha \nP_j+ \rho^i_\alpha \dot{\nA}^\alpha \nP_i-\rho^i_\alpha \nA^\alpha \dot{\nP}_i+\frac{1}{2} \partial_ic^\gamma_{\alpha\beta} \dot{\nX}^i \nA^\alpha \nA^\beta \nB_\gamma \\
             &+c^\gamma_{\alpha\beta} \dot{\nA}^\alpha \nA^\beta \nB_\gamma +\frac{1}{2}c^\gamma_{\alpha\beta} \nA^\alpha \nA^\beta \dot{\nB}_\gamma +\partial_i\widetilde{\rho}^{\alpha j} \dot{\nX}^i \nB_\alpha \nP_j+ \widetilde{\rho}^{\alpha i} \dot{\nB}_\alpha \nP_i-\widetilde{\rho}^{\alpha i} \nB_\alpha \dot{\nP}_i\\
             &+ \frac{1}{2} \partial_i\widetilde{c}_\gamma^{\alpha\beta} \dot{\nX}^i \nB_\alpha \nB_\beta \nA^\gamma +\widetilde{c}_\gamma^{\alpha\beta} \dot{\nB}_\alpha \nB_\beta \nA^\gamma +\frac{1}{2}\widetilde{c}_\gamma^{\alpha\beta} \nB_\alpha \nB_\beta \dot{\nA}^\gamma.\\
      \end{split}
      \end{equation*}

\medskip

Finally, we consider the general procedure of Section \ref{subsub:fixandobs} to relate the expectation values in $\F=Map(T[1]\Sigma, \cZ)$ to those on the reduced theory $\F_{red}=Map(T[1]\Sigma, \cZ_{red})$, having in mind that we already proved $\cZ_{red}\simeq T^*G_A$ in Section \ref{subsec:rlie}.

According to the general treatment, a choice of compatible gauge fixing $\L\hookrightarrow \F$ will locally look like $\L\overset{loc}{=}\L_0 \times \L_{gh}\hookrightarrow\F_0\times\F_{gh}\overset{loc}{=}\F$, with $\L_0 \hookrightarrow Map(T[1]\Sigma,\cZ_{ga=0})$ inducing a lagrangian $\L_{red} \hookrightarrow \F_{red}$ upon reduction by $\fC=Map(T[1]\Sigma, \cC)$ and 
\begin{equation}\label{losev-trick}
\L_{gh}=\{ (\nA^\alpha,\nP_i,\dot{\nX}^i,\dot{\nB}_\alpha)\in \F_{gh}\ |\ \dot{\nX}^i=0,\quad \dot{\nB}_\alpha=0\},
\end{equation}
being defined by setting the fields with $ga<0$ (i.e. the antighosts) to zero. When we restrict the action $S_\F$ to the subset with antighosts set to zero, we obtain
\begin{equation}\label{eff-act-L2}
\begin{split}
S_\F|_{\F_0\times\L_{gh}}=&\int_{T[1]\Sigma}\mu_\Sigma\int_I \Big(\dot{\nP}_id_\Sigma \nX^i+ \nB_\alpha d_\Sigma \dot{\nA}^\alpha\Big)+\Big( -\widetilde{\rho}^{\alpha i} \nB_\alpha \dot{\nP}_i+\frac{1}{2}\widetilde{c}_\gamma^{\alpha\beta} \nB_\alpha \nB_\beta \dot{\nA}^\gamma\Big)+\\
              &+\nP_i\Big(d_I \nX^i+\rho^i_\alpha \dot{\nA}^\alpha\Big)+\nA^\alpha\Big(d_I \nB_\alpha-\rho^i_\alpha  \dot{\nP}_i +c^\gamma_{\beta\alpha} \dot{\nA}^\beta \nB_\gamma \Big).
         \end{split}
         \end{equation}
As recalled from \cite{ale:red} in the general setting of Section \ref{subsub:fixandobs}, we consider an observable $\cO$ on $\F$ which has the $ga$-structure given in eq. \eqref{eq:bfvobs} and induces a reduced observable $\cO_{red}$ on $\F_{red}$. Integrating out the $ga>0$ superfields $\nP_i$ and $\nA^\alpha$ inside of $\langle\cO\rangle_{CSM}$ yields a delta $\delta_{\fC}$ supported on the coisotropic submanifold $\fC=Map(T[1]\Sigma,\cC)$. Following that section further, the resulting integral formally corresponds to $\langle \cO_{red} \rangle$ on $\F_{red}=Map(T[1]\Sigma,\cZ_{red})$. Finally, by Theorem \ref{prop-red}, we have $\cZ_{red}\simeq T^*[1]G_A$ and that $\langle \cO_{red} \rangle$ identifies with the an expectation value for the PSM with target $(G_A,\pi_G)$. This finishes the arguments supporting Claim \ref{main}.

\smallskip

A detailed analysis of how given observables $\cO_{red}$ of the PSM can be homologically represented by observables $\cO$ in the CSM, specializing the general discussion of Section \ref{subsub:fixandobs}, will be detailed elsewhere. The fact that a compatible gauge fixing $\L\hookrightarrow \F$ with the features used above can actually be implemented is a consequence of the discussion in the following Remark, which describes the transverse geometry associated to the quotient $\cC\to T^*[1]G_A$.

\begin{remark}[Algebroid sprays]\label{rmk:spray}
	Following \cite{ale:loc}, one can use a so-called \emph{algebroid spray} $Z\in \mathfrak{X}^1(A)$ for $A$ to construct a local section of the quotient
	$$ q:\cC\to \cZ_{red}\simeq T^*[1]G_A.$$
	Indeed, $Z$ is a vector field whose flow allows one to choose a particular algebroid path $t\mapsto a(t)$ out of each initial condition $a_0\in A$ close enough to the zero section. In this way, one obtains an exponential-type identification 
	$ A \supset U_A \overset{\sim}{\to} U_G \subset G_A, \ a_0 \mapsto [a_t],$
	onto a neighborhood $U_G$ of the identities $1:M\hookrightarrow G_A$ in the groupoid $G_A$; notice that this is global along $M$. This identification can be promoted to one between $T^*[1]U_G$ and $T^*[1]U_A \hookrightarrow T^*[1]PA$ so that (small enough) groupoid elements are associated to the corresponding (small enough) algebroid paths. In this way, we get a local section of the quotient $q$ above,
	$$\cK: T^*[1]U_G\subset T^*[1]G_A \to \cC$$
	defined around the identities in $G_A$. In Section \ref{subsec:Pspray} below, we will explore this construction in detail for the particular case $A=T^*_\pi M$, coming from a Poisson manifold $(M,\pi)$, in which case $Z$ is called a \emph{Poisson spray} and $\pi_G|_{U_G}$ is shown to define a symplectic realization through a specific formula \cite{cra:on}. Finally, we point that $\cK$ can be used to lift ``small" fields from the PSM to the CSM,
	$$ \{\phi \in Map(T[1]\Sigma, T^*[1]G_A): Im(\phi){\scriptstyle \overset{\subset }{}} T^*[1]U_G \}\overset{\cK}{\simeq} Map(T[1]\Sigma, \cC) \subset \underset{\simeq Map(T[1](\Sigma\times I),\M)}{\underbrace{Map(T[1]\Sigma,Map(T[1]I,\M))}}.$$
	Moreover, the restriction to the image of $\cK$ inside $\cC$ can be interpreted as part of the $\L_0$ component of a complete gauge fixing $\L\overset{loc}{=}\L_0\times \L_{gh}$ for the full CSM which leads to an effective PSM gauge fixing $\L_{red}\subset Map(T[1]\Sigma, T^*[1]U_G)$ after reduction, as in the main computation \eqref{eq:maincompu} and Section \ref{subsub:fixandobs}. (In other words, $\cK$ can be used to construct a lift $\L_0$ of a given $\L_{red}$ as described in Remark \ref{rmk:cotangent}.)
\end{remark}

\section{Examples}\label{sec:examples}

This final section is devoted to particularly relevant examples that illustrate how Claim \ref{main} works. In the first simple example, we reprove many of the relevant arguments to gain insight on how they work and to bring concreteness to the abstract treatment of the previous sections. In the last example, taking $A=T^*_\pi M$ induced from a Poisson structure, we make the connection between a partial gauge fixing choice for the CSM and a construction of a symplectic realization in the context of Poisson geometry due to Crainic and Marcut \cite{cra:on}.





\subsection{Lie bialgebras}
In this example, we work out the details of the case in which $M=pt$ is a point and, thus $(A,A^*)=(\g,\g^*)$ is a Lie bialgebra. We also use the example to detail the general constructions. To emphasize the relevant ingredients, we begin with a sub-case in which the dual structure on $\g^*$ is trivial.

\medskip

\noindent{\bf Lie algebra case.} Let $(\g, [\cdot,\cdot])$ be a Lie algebra that we consider it as a Lie bialgebra with $\pi_\g=0$. The Poisson Lie group integrating it is $(G,\pi_G=0)$, the connected and simply connected Lie group integrating $\g$ with the zero Poisson structure. The corresponding double is the semi-direct product $E=\g\ltimes \g^*$ or, in graded terms, given by the internal degree 2 QP-manifold $(T^*[2]\g[1]\cong (\g\oplus\g^*)[1], \omega, \theta=Q_\g+0)$. Hence, Claim \ref{main}  shows the heuristic relation between observables
$$\langle \cO\rangle_{CSM(\Sigma\times I, \g\ltimes\g^*)}=\langle \cO_{red}\rangle_{PSM(\Sigma,(G,\pi_G=0))}.$$

For a 3-dimensional $N$, the space of superfields $\F_N(T^*[2]\g[1])$ for the CSM  is parametrized by two superfields $\bA$ and $\bB$, each superfield has four components $\{\bA_{(q)}, \bB^{(q)}\}_{q=0}^3$ with 
 $$\bA_{(q)}\in\Omega^q(N; \g),\quad \bB^{(q)}\in\Omega^q(N; \g^*) \quad \text{and}\quad |\bA_{(q)}|=|\bB^{(q)}|=q-1.$$ 
 The action  is given by
\begin{equation}
S_{CSM(N,\g\ltimes \g^*)}=\int_{T[1]N}\bB_\alpha d_N \bA^\alpha+\frac{1}{2}c^\gamma_{\alpha\beta}\bA^\alpha\bA^\beta\bB_\gamma.
\end{equation}
Therefore the CSM coincides with the 3d Chern-Simons for the quadratic Lie algebra $\g\ltimes \g^*$ that is also known as $3d$ BF-theory associated to $\g$.

Considering $N=\Sigma\times I$ as before, our boundary conditions reduce to $\bA|_{t=0,1}=0$. Taking the gauge fixing as in \eqref{losev-trick}, we obtain that the action \eqref{eff-act-L2} has the simpler form
\begin{equation}\label{act-liealg}
\begin{split}
(S_{CSM(N,\g\ltimes \g^*)})_{|\F_0\times\L_{gh}}=&\int_{T[1]\Sigma}\mu_\Sigma\int_I  \nB_\alpha d_\Sigma \dot{\nA}^\alpha+
              \nA^\alpha\Big(d_I \nB_\alpha+c^\gamma_{\beta\alpha} \dot{\nA}^\beta \nB_\gamma \Big).
         \end{split}
         \end{equation}
Integrating out the field $\nA^\alpha$ in the path integral underlying $\langle \cO\rangle_{CSM(\Sigma\times I, \g\ltimes\g^*)}$ results in a Dirac delta distribution $\delta_{\frak{C}}$ supported on the coisotropic submanifold  $\frak{C}=\{d_I \nB_\alpha+c^\gamma_{\beta\alpha} \dot{\nA}^\beta \nB_\gamma=0\}$. We recognize this equation as the one given algebroid paths on the Lie algebroid $T^*_{\pi_{\g^*}}[1]\g^*[1]\simeq T^*[1]\g$ associated to the linear Poisson structure $\pi_{\g^*}$ on $\g^*[1]$. Moreover, the remaining term in the action \eqref{act-liealg} is invariant under the infinitesimal gauge transformations
$$\nB_\alpha\to \nB_\alpha+\epsilon(c^\gamma_{\beta\alpha}\nB_\gamma \ell^\beta)\qquad \dot{\nA}^\alpha\to \dot{\nA}^\alpha+\epsilon(d_I\ell^\alpha+ c^\alpha_{\beta\gamma}\dot{\nA}^\beta \ell^\gamma) \quad \text{with}\quad \ell^\alpha\in\Omega^0(I,\g), \ \ell^\alpha|_{t=0,1}=0 .$$
These symmetries restricted to $\cC$ correspond to algebroid homotopies and yield the null directions that must be quotiented out in the coisotropic reduction of Section \ref{subsec:rlie}. The identification 
$$ (\cC/\sim) \simeq T^*[1]G,$$
can be verified directly in this case. If we denote by $q:\cC\to T^*[1]G$ the quotient projection, the induced superfields on the quotient are $\bY^\alpha=q(\dot{\nA}^\alpha), \ \bV_\alpha=q(\nB_\alpha)$ with components $\{\bY^\alpha_{(k)},\bV_\alpha^{(k)}\}_{k=0}^2$ defined by  
$$\bY_{(0)}:\Sigma\to G,\quad \bY_{(k)}\in\Omega^k(\Sigma;\bY_{(0)}^*TG), \quad \bV_{(k)}\in\Omega^q(\Sigma;\bY_{(0)}^*T^*G)$$
and effective action given by
\begin{equation}\label{liealg-eff}
\begin{split}
(S_{CSM(N,\g\ltimes \g^*)})_{eff}=&\int_{T[1]\Sigma} \bV_\alpha d_\Sigma\bY^\alpha=S_{PSM(\Sigma,(G,\pi_G=0))}.
         \end{split}
         \end{equation}

\begin{remark}
In this simple case, the identification $\g \supset U_\g \overset{\sim}{\to} U_G \subset G$ given in Remark \ref{rmk:spray} can be taken to be the ordinary exponential from a neighborhood of zero in the Lie algebra to a neighborhood of the identity in the group. Following that Remark further, we get an induced identification $T^*[1]U_G \simeq T^*[1]U_\g \hookrightarrow T^*[1]P\g$ with which we can relate explicitly the fields of the PSM lying in $U_G\subset G$ with a subset of fields in the CSM described in terms of $\g\oplus \g^*$. Thus, in this simple case, our main claim boils down to the use of exponential coordinates for $G$ and representing elements as certain paths from zero in $\g$. As mentioned in Remark \ref{rmk:spray}, the restrictions on the CSM fields obtained this way can be interpreted as part of a gauge fixing for the full theory.
\end{remark}

As mentioned in Remark \ref{rmk:changeAAast}, the roles of $\g$ and $\g^*$ can be exchanged. In the present simple case, it is equivalent to an integration by parts in \eqref{act-liealg}. The resulting theory corresponds to the following case.

\smallskip

\noindent {\bf Lie coalgebra case.} Let $(\g^*, [\cdot,\cdot]_*)$ be a Lie algebra and consider $(\g, [\cdot,\cdot]=0, \pi_\g)$ the associated Lie bialgebra with trivial Lie bracket. The Poisson Lie group integrating it is $(\g,\pi_{\g})$ where $\g$ is an abelian group under addition and the linear Poisson structure $\pi_{\g}$ is induced by $[\cdot,\cdot]_*$. The corresponding double is the semi-direct product $E=\g\rtimes \g^*$ or, in graded terms, the internal degree 2 QP-manifold $(T^*[2]\g[1]\cong (\g\oplus\g^*)[1], \omega, \theta=0+\pi_{\g^*})$. Hence, Claim \ref{main} gives the heuristic relation between observables
$$\langle \cO\rangle_{CSM(\Sigma\times I, \g\rtimes\g^*)}=\langle \cO_{red}\rangle_{PSM(\Sigma,(\g,\pi_{\g}))}.$$

For the CSM we have the same space of superfields that in the previous case and the action for a $3$ dimensional manifold $N$ is
\begin{equation*}
S_{CSM(N,\g\rtimes \g^*)}=\int_{T[1]N}\bB_\alpha d_N \bA^\alpha+\frac{1}{2}\widetilde{c}_\gamma^{\alpha\beta}\bB_\alpha\bB_\beta\bA^\gamma.
\end{equation*}
Notice that it coincides with the previous action \eqref{act-liealg} once we integrate by parts if $N$ is closed or once we impose the boundary conditions $\{\bA=0\}$ or $\{\bB=0\}$.

By considering $N=\Sigma\times I$ with boundary conditions $\{\bA=0\}$ and making the same gauge fixing as before we arrive to
\begin{equation*}
(S_{CSM(N,\g\rtimes \g^*)})_{|\F_0\times\L_{gh}}=\int_{T[1]\Sigma}\mu_\Sigma\int_I  \nB_\alpha d_\Sigma \dot{\nA}^\alpha+\frac{1}{2}\widetilde{c}_\gamma^{\alpha\beta}\nB_\alpha\nB_\beta\dot{\nA}^\gamma+\nA^\alpha d_I \nB_\alpha.
\end{equation*}
Integrating out the field $\nA^\alpha$ in the path integral underlying $\langle \cO\rangle_{CSM(\Sigma\times I, \g\rtimes\g^*)}$ yields a delta supported on fields $\nB_\alpha$ which are constant along $t\in I$. Hence, making the field redefinition $\bV_\alpha=\nB_\alpha$ and $\bY^\alpha=\int_I \dot{\nA}^\alpha$ we obtain that the
effective action is given by
\begin{equation*}
\begin{split}
(S_{CSM(N,\g\rtimes \g^*)})_{eff}=&\int_{T[1]\Sigma} \bV_\alpha d_\Sigma\bY^\alpha+\frac{1}{2}\widetilde{c}_\gamma^{\alpha\beta\bY^\gamma}\bV_\alpha\bV_\beta\\
=&\int_{T[1]\Sigma} \bV_\alpha d_\Sigma\bY^\alpha+\frac{1}{2}\pi^{\alpha\beta}_{\g^*}\bV_\alpha\bV_\beta=S_{PSM(\Sigma,(\g,\pi_{\g^*}))}.
         \end{split}
\end{equation*}
The above field redefinition was given in \cite{cat:2d} and is a special case of the general coisotropic reduction procedure of Section \ref{subsec:rlie}. Moreover, this case also yields a reduction procedure from the $3d$ BF-theory to the $2d$ BF-theory both associated to the Lie algebra $\g^*$. More general boundary conditions for the Abelian case, i.e. when $[\cdot,\cdot]_*=0$, were intensively studied in \cite{kap:top}. 
     
\smallskip 

\noindent {\bf General Lie bialgebra case.} The combination of the two previous examples yields the Lie bialgebra case $(\mathfrak{g}, [\cdot,\cdot], \pi_\g)$. The corresponding target space of the PSM is given by the Poisson Lie group $(G\rightrightarrows pt, \pi_G)$ integrating this bialgebra. 

Associated to the bialgebra we have its Drinfeld double $E=\g\bowtie \g^*,$ that in graded terms is given by the internal degree $2$ QP-manifold $(T^*[2]\g[1]\cong(\g\oplus\g^*)[1], \omega, \theta=Q_\g\oplus\pi_{\g^*})$. For a surface $\Sigma$, the main Claim \ref{main} gives the following relation between observables
$$\langle \cO\rangle_{CSM(\Sigma\times I, \g\bowtie\g^*)}=\langle \cO_{red}\rangle_{PSM(\Sigma,(G,\pi_G))}.$$
The computations are an straightforward combination of the two previous cases. 

We notice that the $3d$-model coincides with $3d$ Chern-Simons theory for the quadratic Lie algebra given by the double $\g\bowtie \g^*$. However, our result is different from the usual holographic principle, see e.g. \cite{cat:cyl, mnev:hol}, since our 2d theory is again topological instead of conformal. On the other hand, PSM with target a Poisson Lie group are already present in the literature, see e.g. \cite{bon:poi, cal:top}, mainly in connection with coset models $G/G$ as in \cite{ant:top, fal:bou}. Finally, a direct relation between Chern-Simons on $\mathfrak{g}\bowtie\mathfrak{g}^*$ and the PSM on $(G, \pi_G)$, in exactly the same spirit of the present paper, was previously explored by A. Cattaneo and K. Wernli (\cite{cat:priv}). The idea was to try to characterize the quantum group structure associated to $(G, \pi_G)$ via 3d computations on $D\times I$, with $D$ a disk (see also \cite{cat:cyl}).



\subsection{Trivial $A$ case}

This case was explained in \cite{cat:2d} and gives the inspiration for our general statement. Here we compare both approaches.  Let $(A^*, [\cdot,\cdot]_*, \widetilde{\rho})$ be a Lie algebroid and consider $(A, [\cdot,\cdot]=0, \rho_A=0,  \pi_A)$ the associated Lie bialgebroid with trivial Lie bracket and anchor. The Poisson groupoid integrating it is $(A\rightrightarrows M,\pi_{A})$ where $A\rightrightarrows M$ is an abelian groupoid with multiplication given by the fibre-wise addition and the linear Poisson structure $\pi_{A}$ is induced by $[\cdot,\cdot]_*$ and $\widetilde{\rho}$ (as in Section \ref{sec:prelG}). 

The double associated to $(A, A^*)$ is the Courant algebroid $E=A\rtimes A^*$ or, in graded terms, the internal degree 2 QP-manifold $(T^*[2]A[1], \omega, \theta=0+\pi_{A^*})$. Hence, Claim \ref{main} gives the heuristic relation between observables
$$\langle \cO\rangle_{CSM(\Sigma\times I, A\rtimes A^*)}=\langle \cO_{red}\rangle_{PSM(\Sigma,(A,\pi_{A}))}.$$

For the CSM on a $3d$ manifold $N$, the space of superfields $\F=\F_N(T^*[1]A[1])$ is locally given by four superfields $\bX, \bA, \bB, \bP$ each of them with four components living as follows $\bX_{(0)}:N\to M$ and
$$\bX_{(q)}\in\Omega^{q}(N;\bX_{(0)}^*TM),\ \bA_{(q)}\in\Omega^{q}(N;\bX_{(0)}^*A),\ \bB^{(q)} \in\Omega^{q}(N;\bX_{(0)}^*A^*), \ \bP^{(q)}\in\Omega^{q}(N;\bX_{(0)}^*T^*M)$$
of internal degrees  $|\bX_{(q)}|=q, \ |\bA_{(q)}|=|\bB^{(q)}|=q-1,\ |\bP^{(q)}|=q-2$. Thus the action \eqref{Ac-CSM} simplifies to 
\begin{equation*}
S_\F=\int_{T[1]N} \bP_i  d_N\bX^i+ \bB_\alpha d_N\bA^\alpha +\widetilde{\rho}^{\alpha i}(\bX)\bB_\alpha \bP_i+\frac{1}{2}\widetilde{c}^{\alpha\beta}_\gamma(\bX) \bB_\alpha \bB_\beta \bA^\gamma.
\end{equation*}

By considering $N=\Sigma\times I$ with boundary conditions $\{\bA_{|t=0,1}=0\}$ and making the gauge fixing \eqref{losev-trick} we arrive to
\begin{equation*}
S_{\F|\F_0\times\L_{gh}}=\int_{T[1]\Sigma}\mu_{\scriptscriptstyle \Sigma}\int_I \dot{\nP}_id_\Sigma \nX^i+ \nB_\alpha d_\Sigma \dot{\nA}^\alpha -\widetilde{\rho}^{\alpha i} \nB_\alpha \dot{\nP}_i+\frac{1}{2}\widetilde{c}_\gamma^{\alpha\beta} \nB_\alpha \nB_\beta \dot{\nA}^\gamma+\nP_i d_I \nX^i+\nA^\alpha d_I \nB_\alpha.
\end{equation*}
Integrating out the fields $\nA^\alpha, \nP_i$ in the path integral underlying $\langle \cO\rangle_{CSM(\Sigma\times I, A\rtimes A^*)}$ yields a delta supported on fields $\nX^i, \nB_\alpha$ which are constant along $t\in I$. 

In our approach, the effective action is obtained via a coisotropic reduction that can be easily understood as quotient by the symmetries 
$$\dot{\nA}^\alpha\to \dot{\nA}^\alpha+\epsilon d_I \ell^\alpha, \quad \dot{\nP}_i\to\dot{\nP}_i+\epsilon d_I \xi_i, \quad \ell^\alpha\in\Omega^0(I, A),\ \xi_i\in\Omega^1(I, \nX^*T^*M)$$
with  $\ell^\alpha_{|t=0,1}=(\xi_i)_{|t=0,1}=0$. 
Therefore, as done in \cite{cat:2d}, the effective fields can be identified with the following integrals along $I$: $\int_I \dot{\nA}^\alpha, \int_I \dot{\nP}_i$. By direct comparison, the resulting effective action indeed coincides with the PSM on $(A, \pi_A)$.


\subsection{Poisson manifolds, symplectic groupoids and the spray realization}\label{subsec:Pspray}

We begin recalling some general Poisson-geometric facts. 
 We already explained in Example \ref{ex-Lie-alg} and Section \ref{S.bia} that associated to a Poisson manifold $(M, \pi)$ there is a natural Lie bialgebroid $(A=T^*_\pi M,A^*=TM)$. In this case, the associated supergeometric data $(A[1], Q_A, \pi_A)$ has the property that $\pi_A=\omega_{can}^{-1}$ is the canonical (shifted) symplectic structure on $T^*[1]M$ and thus we get an internal degree $1$ QP-manifold $(T^*[1]M, Q_\pi,\omega_{can})$. In contrast with the previous examples, there can be obstructions for the Lie algebroid $A=T^*_\pi M$ to be integrable by a Lie groupoid, see \cite{crfer}. Assuming it is integrable, then the induced Poisson structure $\pi_G$ on $G=G_{T^*_\pi M}\rightrightarrows M$  is actually symplectic $\pi_G=\omega_G^{-1}$. The pair $(G\rightrightarrows M, \omega_G)$ defines a {\bf symplectic groupoid} integrating the Poisson manifold $(M,\pi)$ and the source map $s: (G,\omega_G)\to (M,\pi)$ is a Poisson morphism defining a (strict) {\bf symplectic realization}, see e.g. \cite{cdw}.
 
Let us now move towards the associated field theories.
The corresponding double of $(T^*_\pi M, TM)$ is the Courant algebroid $E=T^*M\bowtie_\pi TM$ defined by 
$$E=TM\oplus T^*M,\quad \langle X+\alpha, Y+\beta\rangle=\alpha(Y)+\beta(X),\quad a(X+\alpha)=X+\pi^\sharp(\alpha), $$
$$\cbrack{X+\alpha}{ Y+\beta}=[X,Y]+[\pi^\sharp(\alpha),Y]-\langle\beta, [\pi,Y]\rangle+[\alpha,\beta]_\pi+L_X\beta-i_Yd\alpha$$
for $X,Y\in\fX(M), \ \alpha,\beta\in\Omega^1(M)$. The supergeometric description of Proposition \ref{sev-roy-corr}  encodes this structure in the internal degree $2$ QP-manifold $(T^*[2]T^*[1]M,\omega, \theta=Q_\pi+\omega_{can}^{-1}).$ 

Our main result, Claim \ref{main}, states that for $\Sigma$ a closed surface we have the relation 
$$\langle \cO\rangle_{CSM(\Sigma\times I, T^*M\bowtie_\pi TM)}=\langle \cO_{red}\rangle_{PSM(\Sigma,(G,\omega_G^{-1}))}.$$
In this case, we shall not repeat all the general arguments. Instead, we shall enrich them by choosing a partial gauge fixing for the CSM using an algebroid spray as in Remark \ref{rmk:spray}. Since $A=T^*_\pi M$, the spray is called a \emph{Poisson spray} $Z\in\fX(T^*M)$ (see below for a definition) and the main result will be that the partially gauge fixed CSM action recovers the formula given in \cite{cra:on} for a symplectic realization $(U\subset T^*M, \omega_Z)$ of $(M,\pi)$. 

\medskip

We start by describing the CSM action for $3d$-manifold $N$. In the remainder of this section, we interchange the notation for variables $a$ and $b$ (with underlying exchange $A=T^*_\pi M \leftrightarrow A^*=TM$, as explained in Remark \ref{rmk:changeAAast}) although we are indeed interested in $G_A$ and not $G_{A^*}$. We count on the reader not getting confused as the role of each field will be explained in detail below. The space of superfields $\F=\F_N(T^*[2]T[1]M)$ is parametrized by $\bX, \bA, \bB, \bP$, each of them with four components, defined as 
$$\bX_{(0)}:N\to M,\quad \bX_{(q)},\bA_{(q)}\in\Omega^{q}(N;\bX_{(0)}^*TM),\quad \bB^{(q)},\bP^{(q)} \in\Omega^{q}(N;\bX_{(0)}^*T^*M)$$
with degrees $|\bX_{(q)}|=q, \ |\bA_{(q)}|=|\bB^{(q)}|=q-1,\ |\bP^{(q)}|=q-2$ and action given by 
\begin{equation}\label{actPoi}
S_{CSM(N,T^*M\bowtie_\pi TM)}=\int_{T[1]N}  \bP_i d_N\bX^i+  \bA^i d_N \bB_i+ \bA^i \bP_i
+\pi^{ij}(\bX)\bB_i \bP_j+\frac{1}{2}\partial_i\pi^{jk}(\bX) \bB_j \bB_k \bA^i.
\end{equation}
Taking $N=\Sigma\times I$, our boundary conditions read $\{\bB|_{t=0,1}=0\}$. We locally split the superfields into the sectors $\F=\F_0\times\F_{gh}$ with $\F_{gh}=\{\nB_i, \nP_i,\dot{\nX}^i, \dot{\nA}^i\}$ and choose the ``UV" gauge fixing $\mathfrak{L}_{gh}=\{\dot{\nX}^i=\dot{\nA}^i=0\}\subset \F_{gh}$. Then the action \eqref{actPoi} reduces to

\begin{equation*}
\begin{split}
S_{CSM(N,T^*M\bowtie_\pi TM)}|_{\F_0\times \L_{gh}}=\int_{T[1]\Sigma}\mu_\Sigma&\int_I  \dot{\nP}_i d_\Sigma \nX^i+\nA^id_\Sigma \dot{\nB}_i-\nA^i \dot{\nP}_i\\
&+\nP_i\Big(d_I\nX^i+\pi^{ji}\dot{\nB}_j\Big)+\nB_i\Big(d_I\nA^i-\pi^{ij}\dot{\nP}_j+\partial_k\pi^{ji}\dot{\nB}_j\nA^k\Big).
\end{split}
\end{equation*}
By integrating out the fields $\nP_i$ and $\nB_i$ in the path integral underlying $\langle \cO\rangle_{CSM(\Sigma\times I, T^*M\bowtie_\pi TM)}$ we obtain deltas supported on the coisotropic submanifold $Map(T[1]\Sigma, \cC)$ locally given by  $\{d_I\nX^i+\pi^{ji}\dot{\nB}_j=0,\ d_I\nA^i-\pi^{ij}\dot{\nP}_j+\partial_k\pi^{ji}\dot{\nB}_j\nA^k=0\}$. 

From the general reduction result in Section \ref{subsec:rlie}, we know that the underlying symplectic quotient for the coisotropic $\cC \subset \cZ= \F_I(\cM)$ yields
\begin{equation}\label{eq:quotsympG} q:\cC\to \frac{\cC}{\sim} \simeq T^*[1]G.\end{equation}
Following Remark \ref{rmk:spray}, the key point in this subsection is to use a Poisson spray to define a (local) section $\cK$ of $q$ on a neighborhood of the form $T^*[1]U_G \subset T^*[1]G$ where $U_G\subset G$ is an open neighborhood of the identity arrows $1_M\subset G$ in $G$. Indeed, the Poisson spray will also allows us to identify $U_G\simeq U\subset T^*M$ with a neightborhood of the zero section on the algebroid $T^*M$. 

\smallskip

We thus proceed to describe Poisson sprays and the induced section $\cK$. A {\bf Poisson spray} for $(M,\pi)$ is a vector field $Z$ on the manifold $T^*M$ satisfying that $$\forall\xi\in T^*M,\quad (d p)_\xi(Z_\xi)=\pi^\sharp(\xi)\quad\text{and}\quad m^*_\lambda Z=\lambda Z \quad \forall\lambda>0,$$
where $p:T^*M\to M$ is the bundle projection and $m_\lambda:T^*M\to T^*M$ is the fibre multiplication by $\lambda\in\bR$. Denoting by $\varphi_t: T^*M\to T^*M$ its flow, the key point is that $t\mapsto \varphi_t(y)$ defines an algebroid path in $A=T^*_\pi M$ for each $y\in T^*M$.
Consider an open neighborhood $U\subseteq T^*M$ of the zero section where the flow of the spray is defined for $t\in [0,1]$. Then, possibly shrinking $U$, the following 2-form
$$\omega_Z=\int_I\varphi_t^*\omega_{can}\ dt$$
is symplectic on $U$. The main statement in \cite{cra:on} says that the bundle projection
$$ p: (U\subset T^*M,\omega_Z)\to (M,\pi)$$
is a Poisson morphism, thus defining a (strict) symplectic realization of $(M,\pi)$. The idea is that, as it will be clear below, $(U,\omega_Z,p)$ is isomorphic to a neighborhood $(U_G,\omega_G,s)$ of $1_M\subset G$ in $G$ with $s$ being the source map of $G$.

Let us now describe the section $\cK$ of the quotient \eqref{eq:quotsympG}. The idea is as follows: recalling that $G$ is identified with algebroid paths in $A=T^*_\pi M$ modulo algebroid homotopies, given $y\in U\subset T^*M$ we can use the flow $\varphi_t$ of the spray to produce an algebroid path
$$ t \mapsto \varphi_t(y).$$
In \cite{crfer}, it is shown that $U$ can be taken small enough so that this procedure defines a map $exp_Z: U \to PA \subset Map(I,T^*M)$ which cuts the algebroid homotopy classes transversally, thus inducing the diffeomorphism $U\simeq U_G$ mentioned above. This diffeomorphism have a natural lift to one between $T^*[1]U$ and a neighborhood of paths inside $\cC$, thus providing the desired section $\cK$ of $q$ in \eqref{eq:quotsympG}. Notice that this construction is global along $M$.

\smallskip

Finally, we use $\cK$ to promote PSM fields into a subset of ``partially gauge fixed" CSM fields, as descrived in Remark \ref{rmk:spray}. Our aim is to compare the corresponding restricted CSM action and the symplectic PSM with target $(U,\omega_Z)$. To that end, we first describe the latter with focus on the restriction to the classical fields $Hom(T\Sigma, T^*U)\subset Map(T[1]\Sigma,T^*[1]U)$ given by vector bundle morphisms\footnote{These are the components of the superfields which have internal degree zero, see e.g. \cite{cat:kon}}. By the general construction recalled on Section \ref{gen}, since $T^*U\simeq TU$ by means of the symplectic structure on $U$, the PSM of $(U,\omega_Z)$ can be defined by fields $Y:\Sigma\to U,\ V\in\Omega^1(\Sigma; Y^*TU).$ Moreover, since $U\subseteq T^*M$, we decompose $Y=(Y_x, Y_b)$ and $V=(V_x,V_b)$ into vertical and horizontal components as
\begin{equation*}
Y_x:\Sigma\to M,\quad Y_b\in\Omega^0(\Sigma, Y_x^*T^*M),\quad V_x\in\Omega^1(\Sigma, Y_x^*TM)\quad \text{and}\quad V_b\in\Omega^1(\Sigma, Y_x^*T^*M).
\end{equation*}
Here, $x$ denote coordinates on $M$ and $b$ fiberwise linear coordinates on $T^*M$.
With these fields and identifications, the classical part of the PSM action \eqref{Act-PSM} becomes
\begin{equation*}
\begin{split}
S^{(cl)}_{\scriptscriptstyle PSM(\Sigma,(U,\omega_Z))}=& \int_{\Sigma}\omega_Z(V^i,d_\Sigma Y^i)+\frac{1}{2}\omega_Z(V^i, V^j)\\
=&\int_{\Sigma}\int_I\Big(\varphi^*_t(\omega_{can})(V^i,d_\Sigma Y^i)+\frac{1}{2}\varphi^*_t(\omega_{can})(V^i, V^j)\Big)dt.
\end{split}
\end{equation*}
On the other hand, we can consider the embedding of (classical) PSM fields into CSM fields induced by $\cK$,
$$ j_Z: Hom(T\Sigma,TU) \to Map(T[1]\Sigma \times T[1]I, T^*[2]T[1]M)=\F$$
which results from the embedding of the internal degree zero part $Hom(T\Sigma,TU)\subset Map(T[1]\Sigma,T[1]U)$, the spray-induced identification $$T[1]U\simeq T^*[1]U \overset{\cK}{\hookrightarrow }\cC \subset Map(T[1]I,T^*[2]T[1]M)$$ and the exponential identification of Section \ref{S3}.

In local coordinates, one characterize the image of $j_Z$ as those fields on $T[1](\Sigma \times I)$ whose $t$-dependence is fixed by the spray flow as follows:
\begin{equation}\label{G1}
\big( X^i(e, t), \dot{B}_i(e, t)\big)=\Big(\varphi^{1,i}_t\big(X(e,0),\dot{B}(e,0)\big),\varphi^{2,i}_t\big(X(e,0),\dot{B}(e,0)\big)\Big)
\end{equation}
for $e\in\Sigma, \ t\in I=[0,1]$, while the other two fields appear as variations of the previous ones
\begin{eqnarray}
A^i(e,t)=\dfrac{\partial\varphi^{1,i}_t}{\partial X^j(e,0)}A^j(e,0)+\dfrac{\partial\varphi^{1,i}_t}{\partial\dot{B}_j(e,0)}\dot{P}^j(e,0)\label{G2}\\
\dot{P}_i(e,t)=\dfrac{\partial\varphi^{2,i}_t}{\partial X^k(e,0)}A^k(e,0)+\dfrac{\partial\varphi^{2,i}_t}{\partial\dot{B}_k(e,0)}\dot{P}^k(e,0)\label{G3}
\end{eqnarray}
where we abbreviated $\varphi^{1,i}_t=\varphi^{1,i}_t\big(X(e,0),\dot{B}(e,0)\big)$ and $\varphi^{2,i}_t=\varphi^{2,i}_t\big(X(e,0),\dot{B}(e,0)\big)$. 

The main result relating this spray-induced partial gauge fixing of the CSM (namely, the image of $j_Z$) to the symplectic realization $\omega_Z$ is the following.

\begin{proposition}
	With the notations above and using the upper-script $(cl)$ to indicate the internal degree zero (i.e. the classical) part of the corresponding action, 
	\begin{equation*}
	S_{CSM(\Sigma\times I, T^*M\bowtie_\pi TM)} \circ j_Z  = 
	S^{(cl)}_{PSM(\Sigma,(U,\omega_Z))}
	\end{equation*}
	where $j_Z:Hom(T\Sigma,TU) \to Map(T[1]\Sigma \times T[1]I, T^*[2]T[1]M)$ is the injection of PSM fields into  CSM fields induced by the Poisson spray $Z\in \fX(T^*M)$ as described using eqs. \eqref{G1}-\eqref{G3} above and $\omega_Z\in \Omega^2(U\subset T^*M)$ is the symplectic realization 2-form defined by $Z$ in \cite{cra:on}.
\end{proposition}


\begin{proof}
	Once we reduce the statement to the coordinate case, we need to show
	$$ \int_{\Sigma}\omega_Z(V^i,d_\Sigma Y^i)+\frac{1}{2}\omega_Z(V^i, V^j) =\int_{\Sigma}\int_I \Big(\dot{P}_i d_\Sigma X^i+A^i d_\Sigma\dot{B}_i-A^i\dot{P}_i\Big)dt,$$
	after the identifications $X^i(e,0)=Y_x^i(e),\ \dot{B}_i(e,0)=Y_b^i(e),\ A^i(e,0)=V_x(e),\ \dot{P}_i(e,0)=V_b^i(e), \ e\in \Sigma$ and where the $t$-dependence of $X,\dot{B}, A, \dot{P}$ is defined by eqs. \eqref{G1}-\eqref{G3}.
	After these identifications, the statement follows from the equalities
	\begin{equation}\label{eq-eq}
	\varphi^*_t(\omega_{can})(V^i,d_\Sigma Y^i)=\dot{P}_id_\Sigma X^i+A^i d_\Sigma\dot{B}_i\quad \text{ and }\quad \frac{1}{2}\varphi^*_t(\omega_{can})(V^i, V^j)=-A^i\dot{P}_i
	\end{equation}
	that we will prove by direct computation.  In order to see these equations observe that
	\begin{equation*}
	\varphi^*_t(\omega_{can})=d\varphi^{1,i}_t\wedge d\varphi^{2,i}_t=\Big(\dfrac{\partial\varphi^{1,i}_t}{\partial Y^j_x(e)}dY^j_x+\dfrac{\partial\varphi^{1,i}_t}{\partial Y^j_b(e)}dY^j_b\Big)\wedge\Big(\dfrac{\partial\varphi^{2,i}_t}{\partial Y^k_x(e)}dY^k_x+\dfrac{\partial\varphi^{2,i}_t}{\partial Y^k_b(e)}dY^k_b\Big)
	\end{equation*}
	and 
	\begin{equation*}
	V^i=V^i_x(e)\frac{\partial}{\partial Y_x^i}+V^i_b(e)\frac{\partial}{\partial Y_b^i}\quad \text{and}\quad d_\Sigma Y^i=d_\Sigma Y^i_x(e)\frac{\partial}{\partial Y_x^i}+d_\Sigma Y^i_b(e)\frac{\partial}{\partial Y_x^i}
	\end{equation*}
	Finally if we combine equations \eqref{G2} and \eqref{G3} with
	\begin{eqnarray*}
	d_\Sigma X^i&=&\dfrac{\partial\varphi^{1,i}_t}{\partial X^j(e,0)}d_\Sigma X^j(e,0)+\dfrac{\partial\varphi^{1,i}_t}{\partial \dot{B}_j(e,0)}d_\Sigma \dot{B}_j(e,0)\\
	d_\Sigma\dot{B}_i&=&\dfrac{\partial\varphi^{2,i}_t}{\partial X^j(e,0)}d_\Sigma X^j(e,0)+\dfrac{\partial\varphi^{2,i}_t}{\partial \dot{B}_j(e,0)}d_\Sigma\dot{B}_j(e,0),
	\end{eqnarray*}
	a further straightforward computation shows that the equations in \eqref{eq-eq} follows as desired.
\end{proof}

\bibliographystyle{plain}
\bibliography{DR}

\end{document}